\documentclass[aps,prd,amsmath,amssymb,footinbib]{revtex4-1}

\usepackage{graphicx}
\usepackage{natbib}

\usepackage{bm}
\usepackage{url}
\usepackage{textcase}
\usepackage{subcaption}
\usepackage{float}
\usepackage{amsthm}
\usepackage{braket}
\usepackage{color}
\usepackage[normalem]{ulem}

\newcommand{\be}{\begin{equation}}
\newcommand{\ee}{\end{equation}}

\begin{document}

\title{Solitonic Boson Stars: Numerical solutions beyond the thin-wall approximation}

\author{Lucas G. Collodel}
 \email{lucas.gardai-collodel@uni-tuebingen.de}

\affiliation{Theoretical Astrophysics, Eberhard Karls University of T\"ubingen, T\"ubingen 72076, Germany}

\author{Daniela D. Doneva}
\email{daniela.doneva@uni-tuebingen.de}
\affiliation{Theoretical Astrophysics, Eberhard Karls University of T\"ubingen, T\"ubingen 72076, Germany}

\begin{abstract}
In this paper we present several set of solutions of static and spherically symmetric solitonic boson stars. Each set is characterized by the value of $\sigma$ that defines the solitonic potential in the complex scalar field theory. The main features peculiar to this potential occur for small values of $\sigma$, but for which the equations become so stiff as to pose numerical challenges. By solving the full system of field equations without simplifications we build the sets for decreasing $\sigma$ values and show how they change their behavior in the parameter space, giving special attention to the region where \emph{thin-wall} configurations dwell. The validity of the thin-wall approximation is explored as well as the possibility of the solution sets being discontinuous. We investigate five different possible definitions of a radius for boson stars and employ them to calculate the compactness of each solution in order to assess how different the outcomes might be. 
\end{abstract}

\maketitle

\section{INTRODUCTION}
\label{sec:intro}
Boson stars are among the most promising candidates for black hole mimickers. Since they first appeared in the literature half a century ago \cite{PhysRev.172.1331,PhysRev.187.1767} as the realization of a free complex scalar field bound by its own gravitational field, many generalizations have been made. Of particular interest is the inclusion of self-interaction potentials that allow bound states to arise even in the absence of gravity, the so called Q-balls \cite{Coleman:1985ki}. These potentials must contain an attractive term while still being bounded from below, such that a degenerate vacuum state occurs with the appearance of a local minimum. The shortest polynomial form of such a potential in a shift-symmetric theory is given by a quadratic, a quartic, and a sixtic term e.g.  $U=c_2\phi^2+c_4\phi^4+c_6\phi^6$, with $c_2, c_6>0$ and $c_4<0$. Q-Balls belong to a particular class called nontopological solitons, whose stability is only possible due to a conserved Noether charge $Q$ associated with the underlying internal symmetry of the theory. Hence the name.

This sixtic potential has been widely investigated in the literature. Static and rotating Q-balls first appeared together in \cite{PhysRevD.66.085003}. Extended solution sets for both Q-balls and boson stars with different gravitational coupling strengths for both static and rotating systems were done in \cite{PhysRevD.72.064002} and extended in \cite{PhysRevD.77.064025} to include odd-parity solutions. Their stability properties were studied via catastrophe theory for both static \cite{PhysRevD.83.044027} and rotating configurations \cite{PhysRevD.85.024045}. Excited solutions of rotating boson stars were studied in \cite{PhysRevD.96.084066}, while rotating solutions of the gauged theory described by this self interaction were given in \cite{PhysRevD.99.104076}.

An interesting particular form of the sixtic potential, mostly referred to as \emph{solitonic}, was proposed in \cite{PhysRevD.13.2739} for solitons in renormalizable theories containing a Hermitian field. Its form is given by $U=\mu^2\phi^2\left(1-\phi^2/\sigma^2\right)^2$ such that the interacting sector is determined by the $\sigma$ parameter. These theories have the property of having a potential that goes to zero at the false vacuum, where $\phi=\sigma$. For large values of $\sigma$ the system resembles that of mini boson stars, but as this parameter's value decreases, the equations of motion get stiffer. In turn, the system has been less investigated in its full nonlinear form, but a special approximation becomes feasible in a certain point of the parameter space where the solution is said to be the \emph{ideal thin-wall configuration}. In this configuration, the scalar field is constant and equal to $\sigma$ within the star and drops off harshly to zero at the star's effective surface transitioning from the false to the true vacuum state. Assuming this configuration is indeed a possible solution of the system, the equations can be simplified considerably and it is within this context that most studies involving solitonic boson stars (SBS) have been made. Two pioneer works in this sense followed one another in the 80's \cite{PhysRevD.35.3637,Friedberg:1986tq}. At the same time, without any kind of approximations, the equations of motion become increasingly stiffer as one approaches the thin-wall regime, making it much more challenging to handle numerically. Note that throughout this paper, the word approximation is meant at the level of the equations. In all numerical schemes there is always some approximation present, such as the choice of discretization for instance.

Boson stars have masses and sizes ranging from atomic to galactic scales, depending on the scalar field's bare mass and the potential defining the theory \cite{Liebling:2012fv}. One of the main motivations to study these systems is the possibility to obtain highly compact solutions of astrophysical relevance whose spacetime serves as stage to phenomenology much similar to that of black holes'. Hence the term black hole mimicker. On the other hand, understanding how to tell compact boson stars from black holes is fundamental to probe their existence. Different types of observations expected from future missions will guide us in this task. Nevertheless, in the context of general relativity alone, the parameter spaces that need to be swept when considering boson stars and hairy black holes \cite{PhysRevLett.112.221101} are rich enough for single observations to map to multiple configurations. Self-gravitating scalar fields alter the spectrum observed from X-ray binaries and the accretion process   \cite{PhysRevD.80.084023,Cao_2016,Meliani_2017,Teodoro_2021,PhysRevD.103.104064,Collodel_2021,Teodoro:2021ezj}, and yield different imaging results than bald black holes \cite{Vincent_2016,Olivares:2018abq} when comparing objects of equal mass and angular momentum. The most stringent constraints will likely come from gravitation waves detection, as they carry imprints from regions beyond the lightring radius, besides providing the clearest evidence of an event horizon, if present. SBS have been considered in some gravitational wave studies. In \cite{PhysRevD.71.044015}, a hybrid scheme is used to calculate the fluxes and waveforms of extreme mass ratio inspirals (EMRI) where the central object is a supermassive SBS in the thin-wall regime. Since then, other works have been done for both EMRIs and similar mass binaries \cite{PhysRevD.88.064046,PhysRevD.94.084031,PhysRevD.95.124005,PhysRevD.96.104058}.

Several boson star solutions containing lightrings are known in different interacting theories but the maximum compactness they might achieve is still an open question, specially given that there is no clear and intuitive definition for it.  
Recently, two papers came out in the literature approaching among other things the question of how compact SBS can actually be. In \cite{Cardoso_2022}, the authors approximated solutions within the thin-wall regime and found that in four spacetime dimensions the maximum compactness is $\mathcal{C}=0.355$. This value is way below the Buchdahl bound ($\mathcal{C}_B=4/9$) \cite{PhysRev.116.1027}, but just above the maximum compactness ($\mathcal{C}_{B+C}=0.354$) a fluid star can have if, besides obeying Buchdal's assumptions, its equation of state is causal \cite{Urbano_2019}. Beyond the approximations supported by the thin-wall regime, the full set of nonlinear equations were solved for different values of the $\sigma$ parameter in \cite{Bo_kovi__2022}, for which the maximum compactness achieved was $\mathcal{C}=0.336$. There, the authors also find good agreement between the solutions obtained numerically and those from a semi-analytical approach and investigate also other forms of self interaction. 

Charged and uncharged SBSs are part of the scope of investigation in \cite{PhysRevD.92.124061}, among with topological and nontopological solitons. The authors find that, for a particular $\sigma$ value, the set of SBSs does not host thin-wall solutions but while the gauged theory (with the appropriately chosen elementary charge) does. As the configurations approach the limit $\phi(0)\rightarrow\sigma$, the mass and total charge diverge, what also occurs to nontopological solitons. Similar results were obtained in \cite{Cardoso_2022}, where the authors showed that new solutions exist for increasing $\phi(0)>\sigma$, departing from the thin-wall regime. The solution curve is then discontinuous and constituted of two disjoint branches.

 In this work, we present several solution families, focusing mainly on low values of $\sigma$ for which the special character of SBS becomes more prominent. In section \ref{sec:theory} we go over the theory of SBS, present the equations of motion, boundary conditions required by the localized regular system, global charges and discuss the validity of the thin-wall approximation. Different possibilities to define an effective radius of a compact object are also discussed. A brief description of the numerical methods employed to solve the system of ODE is given in section \ref{sec:NS} and the results presented in section \ref{sec:sol}. Section \ref{sec:comp} explores the different outcomes and possible pathologist of the previously defined radii in terms of the resulting compactness. We finally conclude in section \ref{sec:conclusions}. Throughout this paper we adopt the metric signature $(-,+,+,+)$ and $c=8\pi G=1$.

\section{THEORY}
\label{sec:theory}
\subsection{Action}
The theory that gives rise to the self-gravitating scalar matter we here present is simply general relativity with a minimally coupled complex scalar field. The action is given by
\be
\label{action}
S=\int \left[\frac{R}{2\kappa}-\partial_\mu\Phi^*\partial^\mu\Phi-U(\vert\Phi\vert)\right]\sqrt{-g}d^4x, 
\ee
where $R$ is the scalar curvature, $\Phi$ is the scalar field with a self-interaction potential $U$, $\kappa$ is the gravitational coupling, $g_{\mu\nu}$ the metric tensor and $g$ its determinant. Varying this action with respect to $g^{\mu\nu}$ leads to Einstein's field equations,
\be
\label{EFE}
R_{\mu\nu}-\frac{1}{2}Rg_{\mu\nu}=\kappa T_{\mu\nu},
\ee
where the energy momentum tensor reads
\begin{align}
T_{\mu\nu}&=g_{\mu\nu}\mathcal{L}_\Phi -2\frac{\partial\mathcal{L}_\Phi}{\partial g^{\mu\nu}} \nonumber \\
&=-g_{\mu\nu}\left[\partial_\alpha\Phi^*\partial^\alpha\Phi+U(\vert\Phi\vert)\right]+2\partial_\mu\Phi^*\partial_\nu\Phi.
\end{align}

The extra equations come from varying eq. (\ref{action}) with respect to the scalar field, yielding the Einstein-Klein-Gordon equation,
\be
\label{KEE}
\left(\Box-\frac{\partial U}{\partial|\Phi|^2}\right)\Phi=0.
\ee

The potential function, together with its parameters, define the complex scalar field theory. The character of the solutions is highly dependent on it and many different kinds have been extensively studied. A particular interesting class of theories is the so called solitonic, in which bound configurations named Q-balls \cite{Coleman:1985ki} exist even in the absence of gravity. In order to achieve this feature, the potential must contain both attractive and repulsive terms, as to profile a local minimum corresponding to the false vacuum state. A well-known form, which we adopt here, is simply
\be
\label{potential}
U(\vert\Phi\vert)=\mu^2\Phi^2\left(1-\frac{\Phi^2}{\sigma^2}\right)^2,
\ee
where $\mu$ is the mass of the boson and $\sigma$ introduces a new energy scale for the interaction. This is a particular case of the more general sixtic potential, as explained in Sec. \ref{sec:intro}, as the two vacuum states are degenerate ($U(|\Phi|=0)=U(|\Phi|=\sigma)=0$). This was first proposed in \cite{Friedberg:1986tq}, but its full numerical investigation without any approximations is somewhat still lacking due to the stiffness properties of the equations of motion that emerge from this choice of potential. Recently, not approximated solutions have been reported in \cite{Cardoso_2022} for a gauged theory and for a set of $\sigma$ values in the not gauged case in \cite{Bo_kovi__2022}. Here, however, we intend to find extensive sets of solutions for low values of $\sigma$. Throughout the rest of this paper we assume $\kappa=1$.

\subsection{Ans\"{a}tze and Equations}
The objects we describe produce a static and spherically symmetry spacetime which is asymptotically flat. We employ spherical coordinates such that the line element is written as
\be
\label{ds2}
ds^2=-e^\nu dt^2+e^\lambda dr^2+r^2d\Omega^2,
\ee
and $\nu$ and $\lambda$ are the only two unknown metric functions.

The scalar field is time dependent, but only harmonically such that its Lagrangian remains stationary. Thus, 
\be
\label{ansatz_field}
\Phi(t,r)\equiv \phi(r)e^{i\omega_st},
\ee
where $\omega_s$ is the field's natural frequency. Plugging these parametrizations into eq. (\ref{EFE}) and (\ref{KEE}) results in a set of three ODE for the functions $\nu$, $\lambda$ and $\phi$ that we need to solve. We cast them as
\be
\label{eql}
\lambda'+\frac{e^\lambda-1}{r}-r\phi'^2-re^\lambda U-re^{\lambda-\nu}\omega_s^2\phi^2=0,
\ee
\be
\label{eqf}
\nu'+\frac{1-e^\lambda}{r}-r\phi'^2+re^\lambda U-re^{\lambda-\nu}\omega_s^2\phi^2=0,
\ee
\be
\label{eqphi}
\phi''+\left(\frac{\nu'-\lambda'}{2}+\frac{2}{r}\right)\phi'+e^{\lambda-\nu}\omega_s^2\phi-\frac{1}{2}e^\lambda\frac{\partial U}{\partial\phi}=0.
\ee

\subsection{Boundary Conditions and Asymptotic Expansions}

The set of equations (\ref{eql})-(\ref{eqphi}) require us to prescribe boundary conditions as to grant regularity everywhere and asymptotic flatness at spatial infinity. To that end, we have at the origin
\be
\label{BC:0}
\lambda(0)=0, \qquad \phi'(0)=0,
\ee
while at infinity the scalar field must trivialize
\be
\label{BC:1}
\lambda(r\sim\infty)=0, \qquad \nu(r\sim\infty)=0, \qquad \phi(r\sim\infty)=0.
\ee

Near the center of the star, by expanding each function in a polynomial series and solving the equations for each power of $r$, we find the leading order contributions to be

\be
\label{AEphi}
\phi(r\sim 0)=\phi_0+\frac{\phi_0\left[(3\phi_0^2-\sigma^2)(\phi_0^2-\sigma^2)e^{\nu_0}-\omega_s^2\sigma^4\right]}{6e^{\nu_0}\sigma^4}\delta r^2,
\ee

\be
\label{AEf}
e^{\nu(r\sim 0)}=e^{\nu_0}-\frac{\phi_0^2\left[(\phi_0^2-\sigma^2)^2e^{\nu_0}-2\omega_s^2\sigma^4\right]}{3\sigma^4}\delta r^2,
\ee

\be
\label{AEl}
e^{\lambda(r\sim 0)}=1+\frac{\phi_0^2\left[(\phi_0^2-\sigma^2)^2e^{\nu_0}+\omega_s^2\sigma^4\right]}{3e^{\nu_0}\sigma^4}\delta r^2,
\ee

while at infinity

\be
\label{AEIphi}
\phi(r\rightarrow\infty)=c_0\frac{\exp\left(-r\sqrt{\mu^2-\omega_s^2}\right)}{r},
\ee

\be
\label{AEIfl}
e^{-\lambda(r\rightarrow\infty)}=e^{\nu(r\rightarrow\infty)}=1-\frac{2M}{r}+\mathcal{O}(r^{-2}),
\ee
where $M$ is the ADM mass of the star.

\subsection{Charge, Mass and Radius}
\label{Sub:Charge}
The complex scalar field theory is endowed with a global $U(1)$ symmetry, such that its Lagrangian is invariant under transformations of the kind $\Phi\rightarrow\Phi e^{i\alpha}$, for any constant $\alpha$. This gives rise to a conserved Noether current,
\be
\label{eq:NoetherCurrent}
j^\mu=-i\left(\Phi^*\partial^\mu\Phi-\Phi\partial^\mu\Phi^*\right).
\ee

The projection of this current onto the future-directed unit timelike vector $n^\mu=(e^{-\nu/2},0,0,0)$ gives a charge density such as measured by a fiducial observer. Hence, the integral of this density over the whole three spatial volume $\Sigma$ yields a conserved Noether charge, associated with the particle number of the bosonic ensemble,

\be
\label{NoetherCharge}
Q=\int_\Sigma j_\mu n^\mu dV,
\ee
whence it is clear that $\partial_t\Phi=0\rightarrow Q=0$.

The global mass of the system can be extracted asymptotically with aid of eq. (\ref{AEIfl}). Due to asymptotic flatness, this matches the Komar integral in terms of the Killing vector associated with stationarity, $\eta^\mu=(1,0,0,0)$,
\be
\label{KomarMass}
M=-2\int_\Sigma R_{\mu\nu}n^\mu\eta^\nu dV=-\int\left(T^t{}_t-\frac{1}{2}T\right)\sqrt{-g}dr.
\ee

It is useful to define a Komar density, given by the integrand of the above expression, namely $\rho_K=\left(\frac{1}{2}T-T^t{}_t\right)\sqrt{-g}$. Similarly, integrating eq. (\ref{eql}) by parts and using the expansion (\ref{AEIfl}), one finds
\be
\label{IntMass}
M=-\frac{1}{2}\int r^2T^t{}_t dr,
\ee
such that another density, as measured by a fiducial observer, can be defined $\rho_F=-\frac{1}{2}r^2T^t{}_t$.

Boson stars possess no surface, i.e. there's no point where the density goes abruptly to zero as the pressure reaches the same value. The scalar field extends to infinity, but falls off exponentially as given by eq. (\ref{AEIphi}). The solitonic potential (\ref{potential}) allows for solutions characterized by a sudden drop of the scalar field similar to a Heaviside function as discussed bellow, but even then there is always a residual tail in its profile. Hence, there is no unique and straightforward definition of a radius, and consequently, of the compactness of a boson star. Mostly, the radius is assumed to be that which defines a surface that envelops some percentage of total mass or charge of the star. Since these stars are spherically symmetric and because of Birkhoff's theorem, it is also interesting to define a radius as the point where some geometrical quantity matches that of Schwarzschild to a chosen degree of accuracy. In what follows, we make five different definitions of radius:
\begin{itemize}
\item $R_1$: The radius of the 2-Sphere that envelops $99.9\%$ of the mass, as the integral of the Komar density.
\item $R_2$: The radius of the 2-Sphere that envelops $99.9\%$ of the mass, as the integral of the fiducial density.
\item $R_3$: The radius of the 2-Sphere where $g_{tt}$ matches Schwarzschild to $0.999$ accuracy.
\item $R_4$: The radius of the 2-Sphere where $g_{rr}$ matches Schwarzschild to $0.999$ accuracy.
\item $R_5$: The radius of the 2-Sphere where the Kretschmann scalar $\mathcal{K}\equiv R_{\mu\nu\gamma\tau}R^{\mu\nu\gamma\tau}$ matches Schwarzschild to $0.999$ accuracy.
\end{itemize} 

The compactness is defined as the ratio between the total mass and the radius and thus, comprehends also five definitions such as $C_i\equiv M/R_i$ for $i\in[1,5]$.
\subsection{Thin-Wall Regime}

Ideally, thin-wall configurations consist of solutions whose scalar field profiles as a Heaviside function, being constant in the interior region such that $\phi=\phi_0$ for $r\in[0,R_S]$ and zero thereafter, $\phi=0$ for $r>R_s$. The point where the scalar field drops abruptly to zero, $R_S$ is then identified with the star's surface and all the solitonic particles pile up there such that the interior is empty. Due to spherical symmetry and the shell theorem, spacetime is then flat inside and the metric function $g^{rr}$ is discontinuous at $R_S$, where it jumps from one to $\log\left(1-2M/R_S\right)$. According to equations (\ref{eql})-(\ref{eqphi}), this regime is only exact if either the field is trivial (and spacetime is flat everywhere), or if $\phi=\sigma$ in the interior and $\omega_s=0$. Recall that the latter corresponds to the system being in a false vacuum state, $U=\partial_\phi U=0$. However, a trivial field's frequency yields zero Noether charge and therefore the regime can only exist approximately. 

In fact, we can expand eq. (\ref{eqphi}) around $\phi=\sigma+\delta\phi$ at an arbitrary location inside the star. Assuming $\delta\phi\ll\sigma$ and since $\delta\phi'$ and $\delta\phi''$ are higher order contributions, one arrives at 
\be
\label{AESp}
\delta\phi\sim\frac{\omega_s^2\sigma}{4e^\nu-\omega_s^2},
\ee 
in agreement with the statement above. Indeed one can substitute $\phi=\sigma+\delta\phi$ from the equation above in eq. (\ref{AEphi}) to verify that the second order contribution trivializes at the center. In fact, there are many solutions which resemble the ideal thin-wall configuration. Those solutions comprehend the \emph{thin-wall regime}, where the expansion above is valid. In this domain, as $\delta\phi$ decreases, the stiffer the equations of motion become due to the loss of smoothness in both $\phi$ and $\lambda$.


\section{Numerical Scheme}
\label{sec:NS}
In order to find the solutions we make no assumptions of their nature, i.e. we do not employ usual approximations describing system within the thin-wall regime, but rather solve the full set of nonlinear equations and infer from the results where this region resides.
The set of ODE eqs. (\ref{eql})-(\ref{eqphi}) is solved by employing a B-spline collocation method with a relaxed Newton scheme for the nonlinear system with aid of the BVP solver Colsys \cite{10.1145/355945.355951}. In every case, we adopted a tolerance of $10^{-7}$ for each component for the Newton scheme. Einstein's field equations yield a fourth linearly independent equation, which is computed as a consistency check. The solution is accepted if the norm of the residual averaged by the number of internal grid points is below our tolerance.

Throughout the parameter space we adopt two strategies to find the solutions. In the regions where the central value of the scalar field varies monotonically from one solution to the next, we promote $\omega_s$ to a variable by adding to the set the simple ODE $d\omega_s/dr=0$. By doing so, we suggest a value of this parameter as an initial guess, but gain one extra boundary condition to assign. Therefore, we set the two BCs in eq. (\ref{BC:0}), the last two in eq. (\ref{BC:1}) and also set the central value of the scalar field $\phi(0)=\phi_0$. Each solution serves well as an initial guess to the next one until we reach a turning point in $\phi_0$. In this region, we assign values for $\omega_s$ instead and let $\phi_0$ be fixed by the solution. In order to obtain solutions covering the whole space, we compactify the radial coordinate as $r=x/(1-x)$, and work with $x$ instead, solving in the interval $x\in[0,1]$ corresponding to $r\in[0,\infty)$.

The main aim of this work is to evaluate and assess entire families of solutions characterized by a particular value of $\sigma$. As this parameter gets smaller, more and more solutions fall within the thin-wall regime, and smaller is the minimum value of $\delta\phi$ from eq. (\ref{AESp}) therein. Thus, the system becomes gradually stiffer until a point where our numerical methods fail to converge within reasonable tolerance. As we keep increasing $\phi_0$ convergence is eventually regained and new solutions are found in a different part of the parameter space. In turn, we obtain two disjoint branches of solutions. It is not possible to assert whether solutions do indeed exist everywhere between the branches for every value of $\sigma$. Shooting methods with a variety of algorithms suitable for stiff problems were used with arbitrary precision and they all rendered instabilities. More sophisticated methods are required to evaluate this matter. Therefore, we present solution sets with decreasing $\sigma$ only until the first one to present a gap, i.e. a region in the thin-wall regime where no solutions were found. A similar discontinuity was reported in \cite{Cardoso_2022} for gauged solitonic stars with sufficiently large elementary charge, in the same region where the thin-wall regime occurs, but for considerably larger $\sigma$.

\section{Solutions}
\label{sec:sol}
\begin{figure}
\centering
\includegraphics[width=0.48\linewidth]{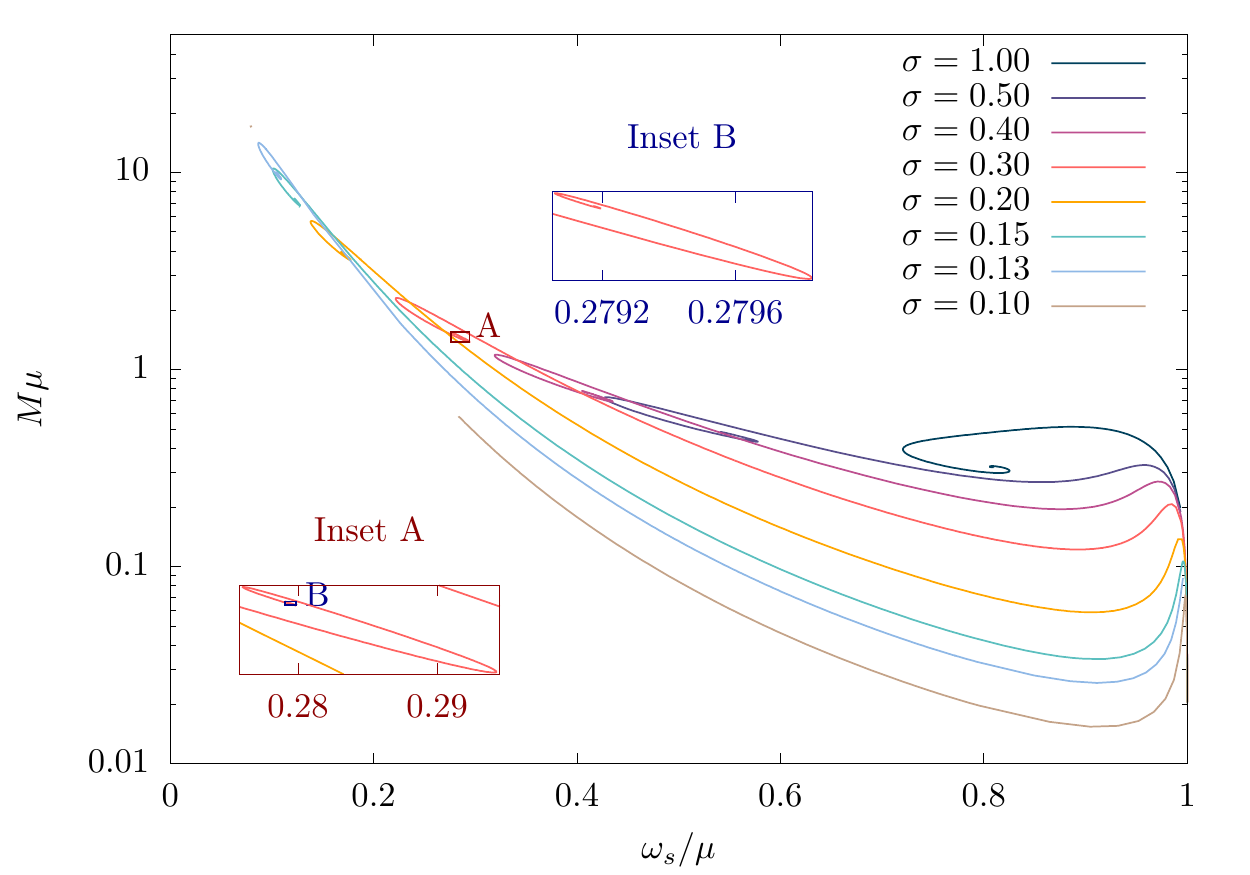}
\includegraphics[width=0.48\linewidth]{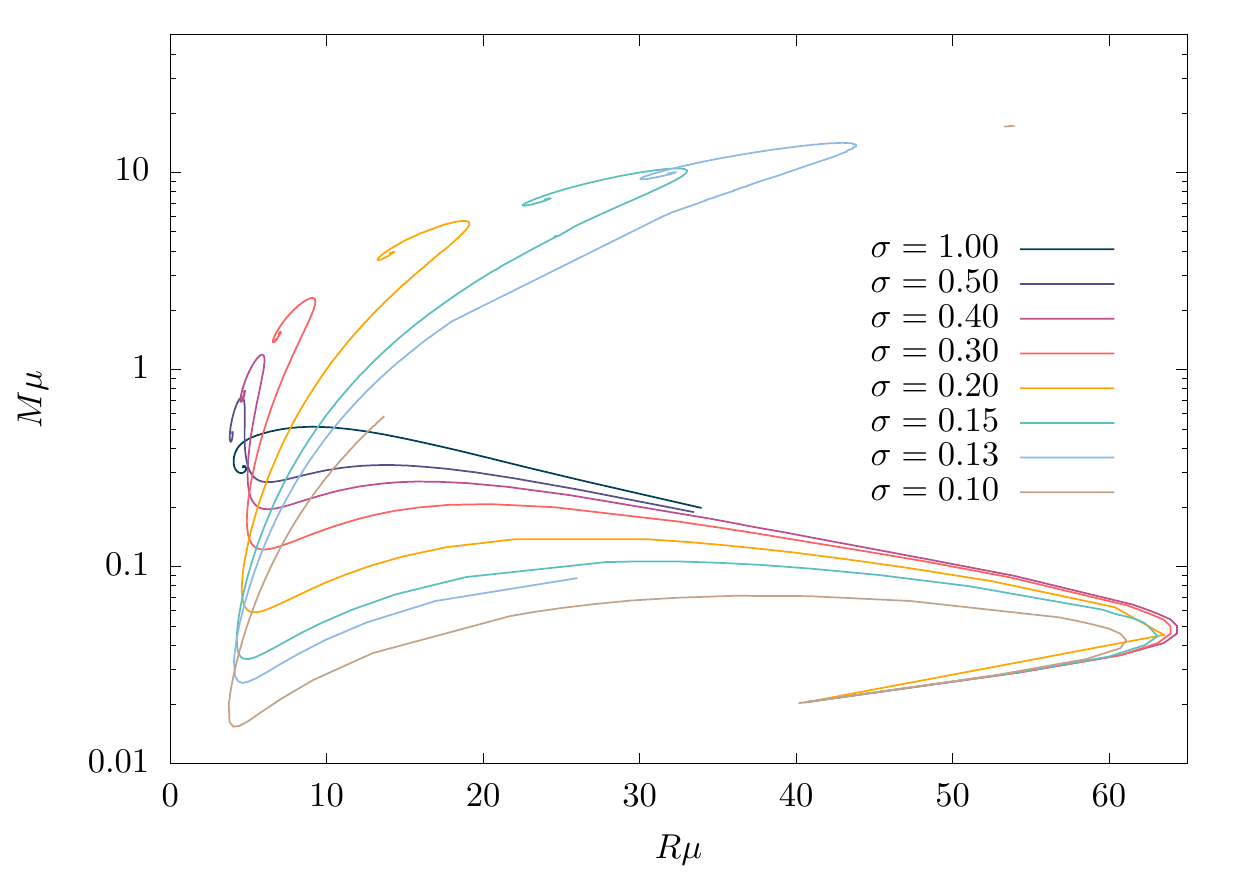}
\caption{Solution sets for eight values of $\sigma$. \emph{Left:} Mass vs. frequency diagram. Similarly to the free theory and to other interacting potentials, the curves wind up around a fixed point in $\omega_s$. The $\sigma=1.0$ case resembles the sets of mini boson stars where the upper branch (before the turning point at $\min\omega_s$) contains only the global maximum in the mass. For the other sets, there is a local maximum of the mass which approaches the limiting solution at $\omega_s\sim 1$ the more $\sigma$ decreases, while the global maximum is precisely at $\min\omega_s$. Two insets are present (case $\sigma=0.3$) to highlight the spiral trait in the parameter space. \emph{Right:} Mass vs. radius diagram. Lower values of the interacting parameter yield sets of solutions containing more compact solutions.  }
\label{Fig:0}
\end{figure}

We have computed extensive sets of solutions for eight different values of $\sigma$, namely $\sigma=\{0.1, 0.13, 0.15, 0.2, 0.3, 0.4, 0.5, 1.0\}$. We recall that the limit $\sigma\rightarrow\infty$ corresponds to the free theory that gives rise to mini boson stars.  Fig \ref{Fig:0} displays the $\omega_s$ vs. $M$ diagram on the left panel and $R$ vs. $M$ on the right one, for the definition of the radius given in terms of the Kretschmann scalar ($R_5$). The usual spiraling behavior peculiar to solitonic stars is also present in this diagrams. This is highlighted by the insets shown in the left panel, and solutions seem to exist indefinitely with increasing $\phi_0$ and converging $\omega_s$.  Furthermore, the smaller the value of $\sigma$, the greater the range of solutions in $\omega_s$, $M$ and $R$, and more compact solutions appear. The $\sigma=1.0$ case resembles much more the mini boson star, as there is only a global maximum of the mass before the turning point on the left panel, indicating that this parameter value is already too large to unveil the distinct features brought up by the interacting theory. Note that the curve for the lowest value $\sigma=0.1$ is composed by two disjoint branches. The solutions of the second branch lie too near on another in these diagrams but will become clearer below. Nevertheless, it is visible on the right panel around $R\sim 50$ and $M\sim 17$. No solutions were found between those branches and the gap becomes even greater for lower values of the self-interaction parameter, for which reason we take this to be our limiting solution set. Similar values might appear for differently scaled Lagrangian densities, which in reality rescales the value of the field and the parameters involved. Recently, an involved study was done in \cite{Bo_kovi__2022}, for both Q-balls and boson stars, where the authors compare numerical results with analytical approximation made by considering three different zones in the soliton, namely the interior, the boundary and the exterior. In two of their sets the $\sigma$ value is small enough to host thin-wall solutions. Nevertheless, to our knowledge, thorough investigation of lengthy sets of solutions for such small values of $\sigma$ as well as a detailed analysis of the possible (non)existence of solutions in this regime has not been reported in the literature before without imposing approximations.

\begin{figure}
\centering
\includegraphics[width=0.48\linewidth]{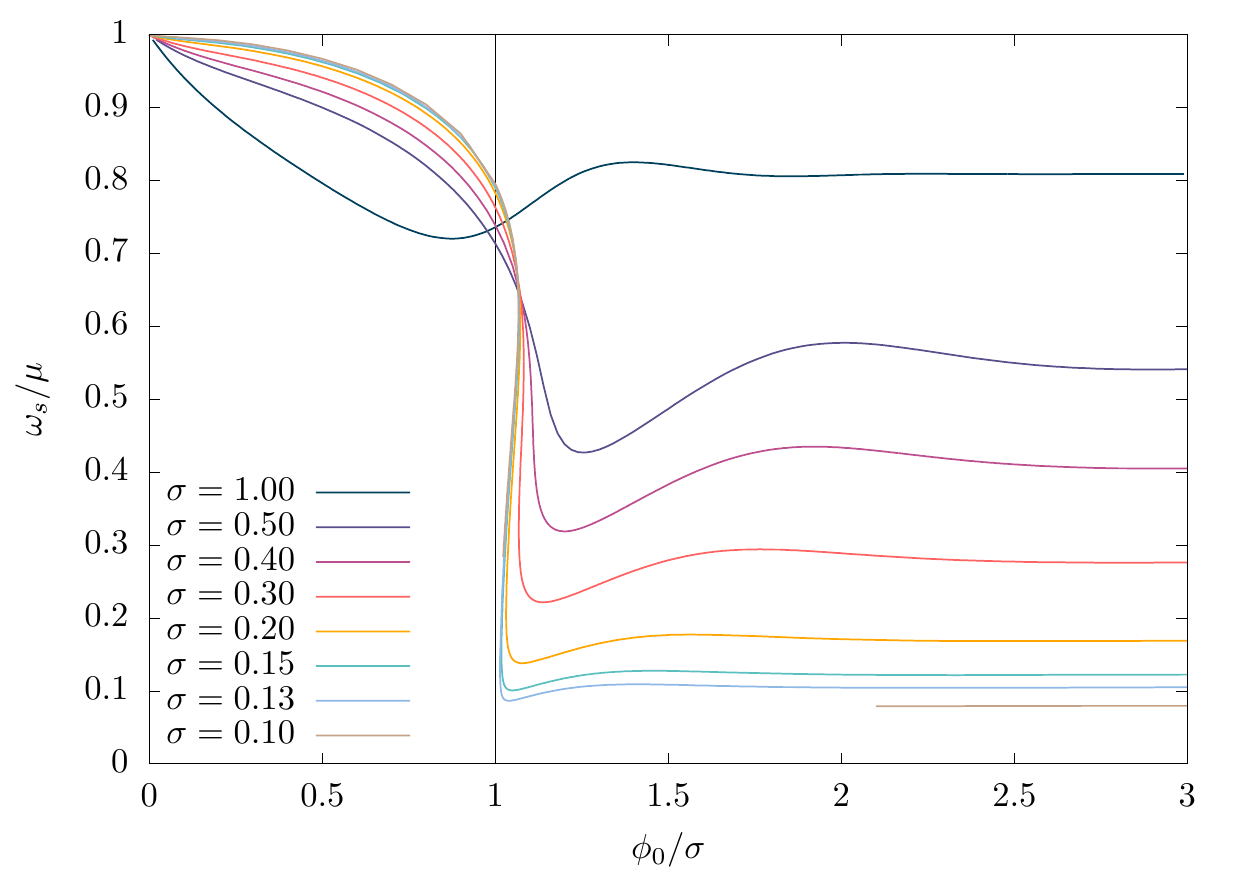}
\includegraphics[width=0.48\linewidth]{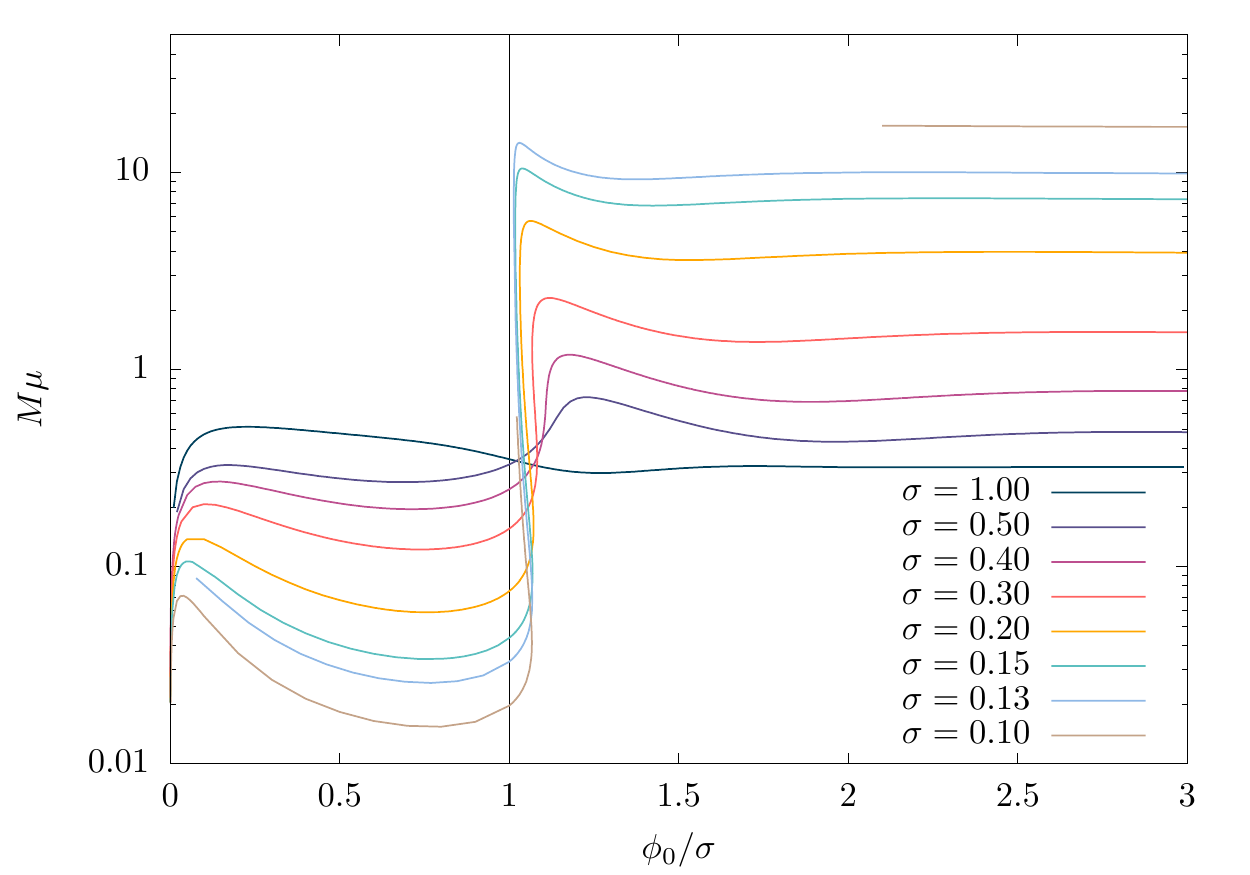}
\caption{Solution sets for eight values of $\sigma$. The set $\sigma=0.1$ features two discontinuous branches as solutions could not be found in the region in between. \emph{Left:} The field's frequency against the central value of the field. We notice a backbend in the family of solutions with $\sigma=0.3$ or lower, meaning that the central value does not uniquely determine the solution. This backbend branch where $\omega_s$ decreases with slightly decreasing $\phi_0$ is where the thin-wall solutions dwell. The smaller the self-interaction parameter is, the closer the curve gets to the vertical line which characterizes solutions with $\phi_0=\sigma$, but also the greater the gap between the branches become. 
\emph{Right:} Mass with respect to the central value of the field for the same families of solutions. Decreasing $\sigma$ increases the maximum of the mass and overall the solutions here presented span over two order of magnitude in the mass.}
\label{Fig:1}
\end{figure}

In Fig. \ref{Fig:1} we present the parameter space of the solution sets, as well as the mass vs. the field's central value ($\phi_0$) diagram. While for large enough values of $\sigma$ the solutions can be parametrized by $\phi_0$, as mini boson stars can, this fails to be true for lower values as the solutions are not uniquely determined by it. Instead, by following the curves from the top left ($\omega_s/\mu=1, \phi_0/\sigma=0$), as they cross the vertical line corresponding to $\phi_0/\sigma=1$ they bend back towards this line and this effect is enhanced with decreasing $\sigma$. After reaching a local minimum of $\phi_0>\sigma$, the central value of the scalar field starts to grow back and from there on increases monotonically while slightly oscillating around a particular value of $\omega_s$, which translates to the spiraling regions of Fig. \ref{Fig:0}. It is precisely at this backbend region that solutions in the thin-wall regime dwell. The vertical line illustrates where solutions characterized by their cores being in false vacuum state would lie. Because of eq. (\ref{AESp}), we know that the curves cannot cross this line after the backbend, as the ideal thin-wall solution would be the point ($\phi_0=\sigma, \omega_s=0$) which is not a solution. 

The solution set of $\sigma=0.1$ is more visible in this figure as the second branch spans a wider range in $\phi_0$  than in the previously shown parameters. The set presents a maximum mass of $M_\text{max}=17.3/\mu$ and a minimum of the frequency of $\omega_{s\text{min}}=0.079\mu$. However, if solutions do exist in the gap, meaning they were not found due to numerical challenges, one can expect from the behavior of the other sets that the maximum of the mass is to be higher than this value and similarly the minimum of the frequency smaller than the one found.  In fact, discontinuous sets were found for gauged solitonic stars if the elementary charge is large enough \cite{PhysRevD.92.124061,Cardoso_2022}. The breaking point also happens as the thin-wall solutions approach the $\phi_0=\sigma$ limit. As reported in those papers, the mass seems to diverge at this point and $g_{tt}\rightarrow0$. In contrast, we will see below that for SBSs $g_{tt}$ tends to zero at the center of the spiriling region shown in \ref{Fig:0}, as $\phi_0$ keeps increasing.

\begin{figure}
\centering
\includegraphics[width=0.48\linewidth]{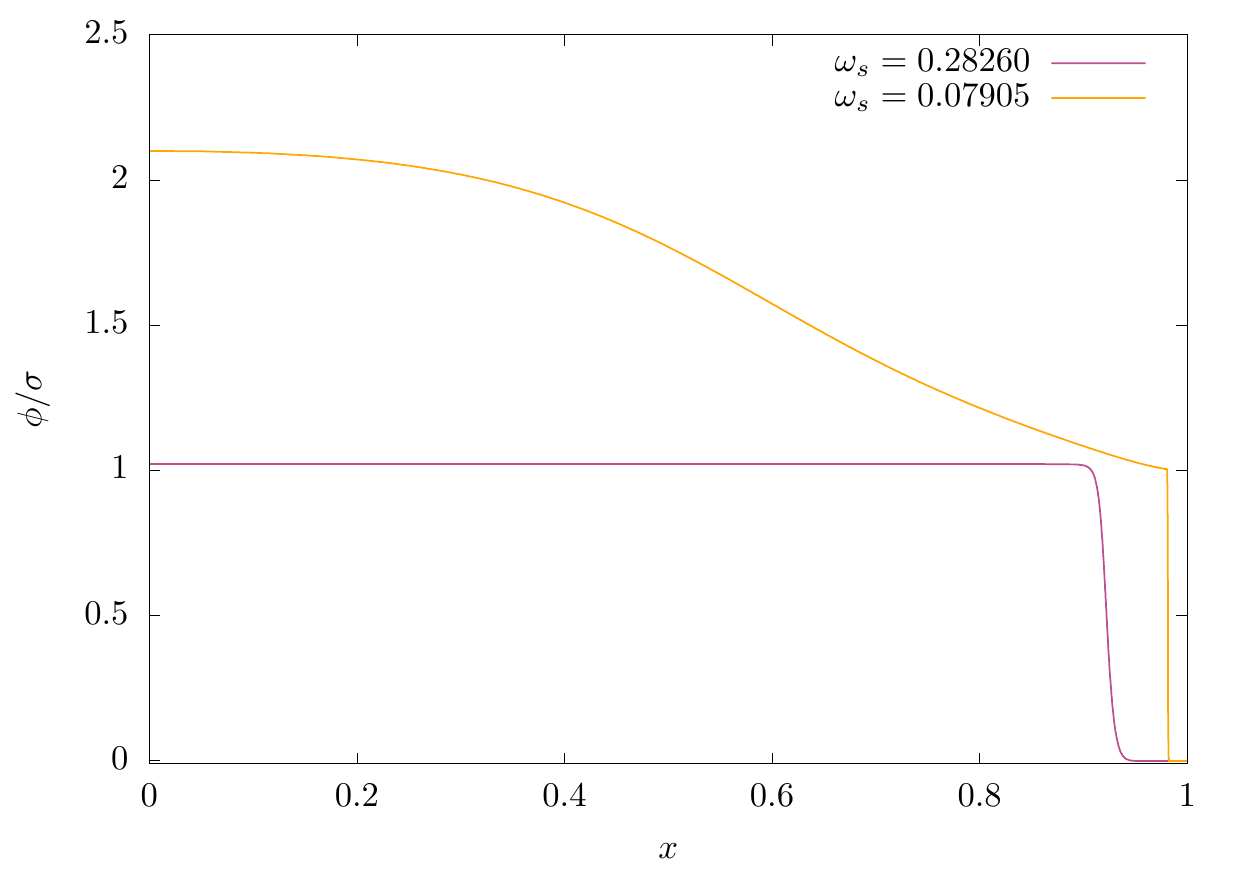}
\includegraphics[width=0.48\linewidth]{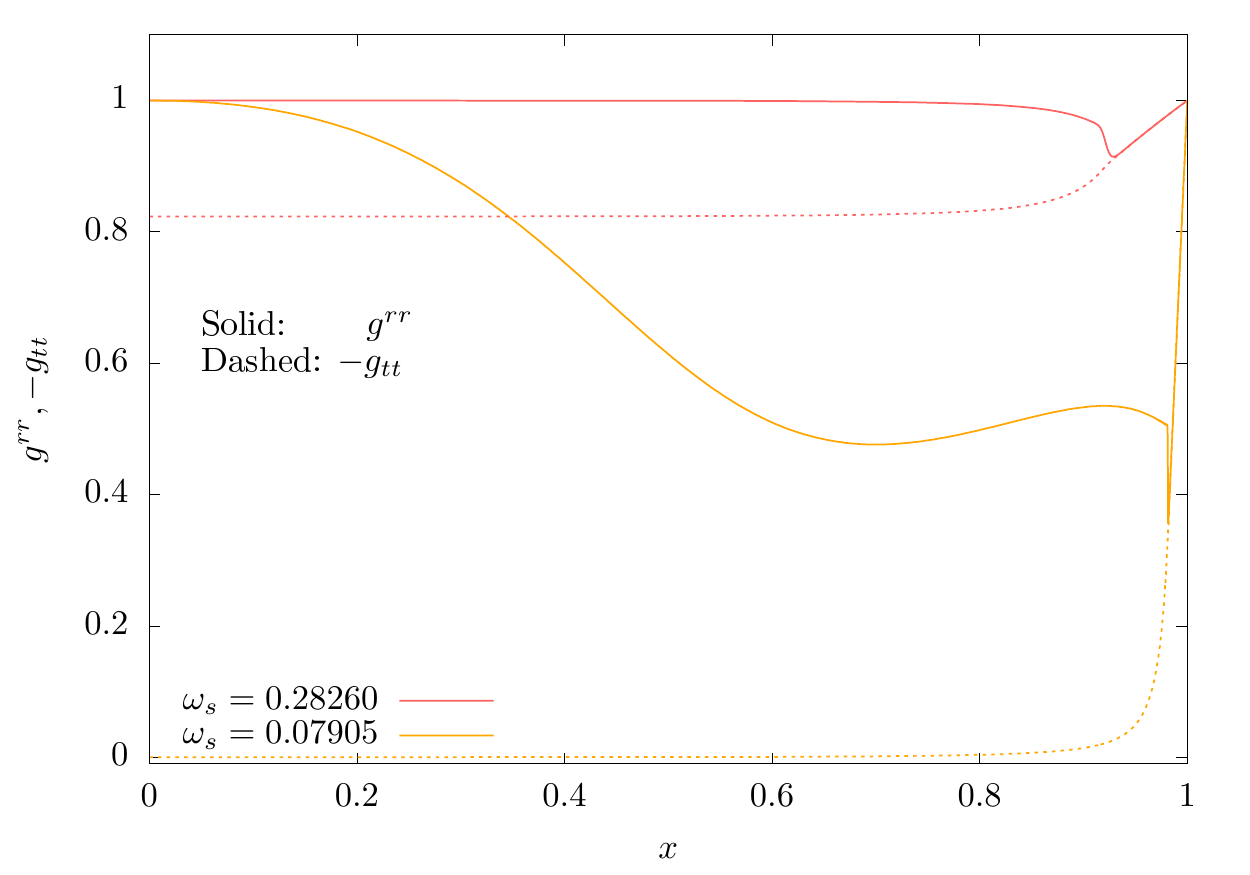}
\caption{Limiting solutions at the gap of the set $\sigma=0.1$. \emph{Left:} Scalar field profile with respect to the compactified coordinate $x$. The system right before the gap falls within the thin-wall regime, i.e. nearly constant $\phi$ in the interior followed by an abrupt drop off to zero at the effective surface. At the other end of the gap, where solutions were again found, the system has departed considerably from this regime, but there clearly is a similar drop off for the field.
\emph{Right:} The metric components $g_{tt}$ and $g^{rr}$ for both solutions. In the interior of the first solution these components are almost constant, yielding an almost flat spacetime. That is not the case for the second solution, for which $g^{rr}$ is much less smooth near the effective surface and $g_{tt}$ gets very close to zero in the interior.}
\label{Fig:2}
\end{figure}

In Fig. \ref{Fig:2} we show the behavior of the scalar field and the metric functions for the two limiting solutions at the edges of the gap in the set of $\sigma=0.1$. Notice that $x$ is the compactified coordinate defined previously. In both cases, the scalar field drops to zero very abruptly but the interior behaviors are fairly different. In the case depicting the first edge of the gap, the solution is in the region of the thin-wall regime. The central value is very close to $\sigma$ and the first and second derivatives of the field are negligible until the star's surface. The second edge solution is different as the field is still dynamical in the interior. This is reflected in the profile of the metric functions, shown in the right panel. The thin-wall solution presents almost constant metric functions in the interior, meaning that the spacetime is nearly flat inside the star. This is simple to conceive since the thin-wall star can be thought of as a massive shell of scalar matter and because of spherical symmetry any point inside this shell is not gravitationally affected by it. If the field keeps varying in the interior, as it is for the second solution, the metric functions  must of course behave differently. Note that $-g_{tt}$ drops very rapidly as one moves closer to the center of the star where $\exp\nu_0\sim 1.5\times 10^{-4}$.

\begin{figure}
\centering
\includegraphics[width=0.48\linewidth]{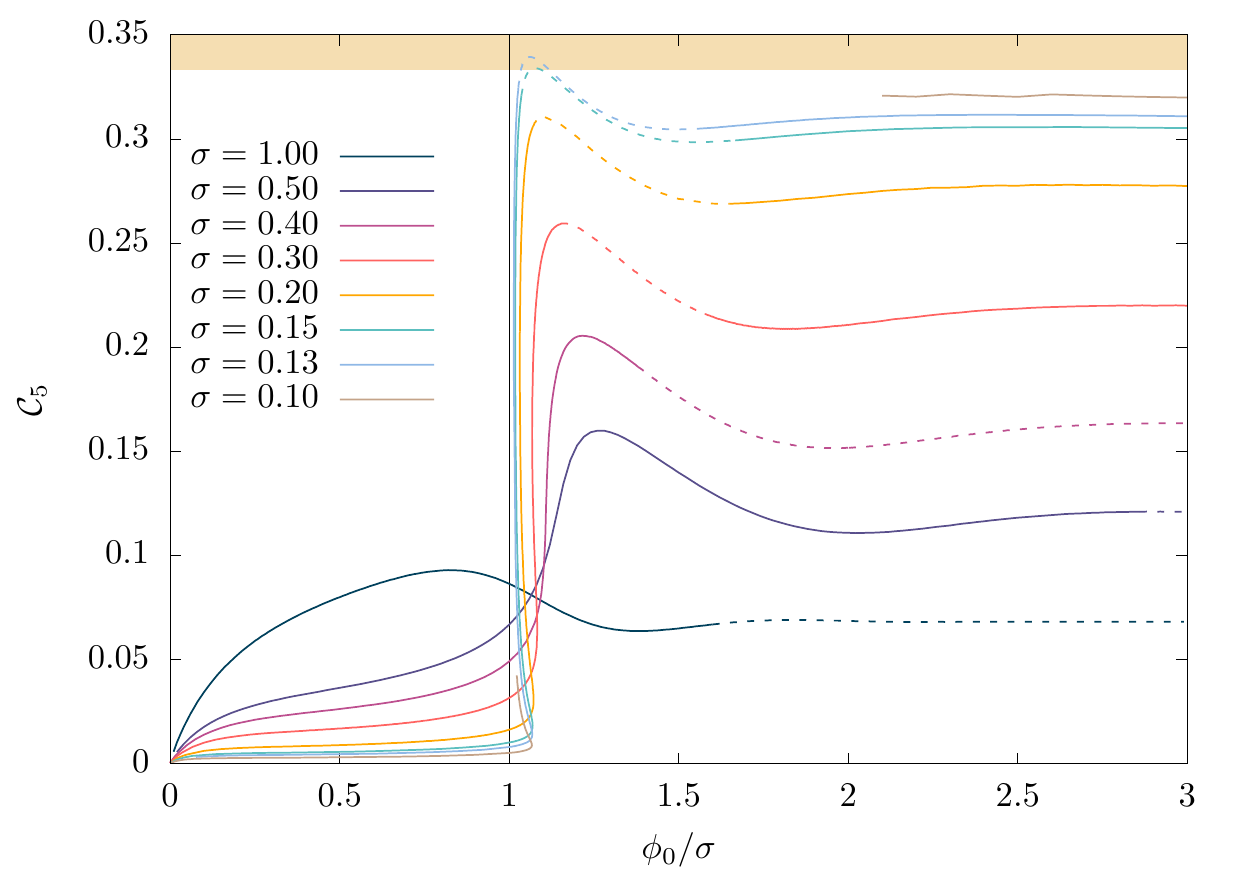}
\includegraphics[width=0.48\linewidth]{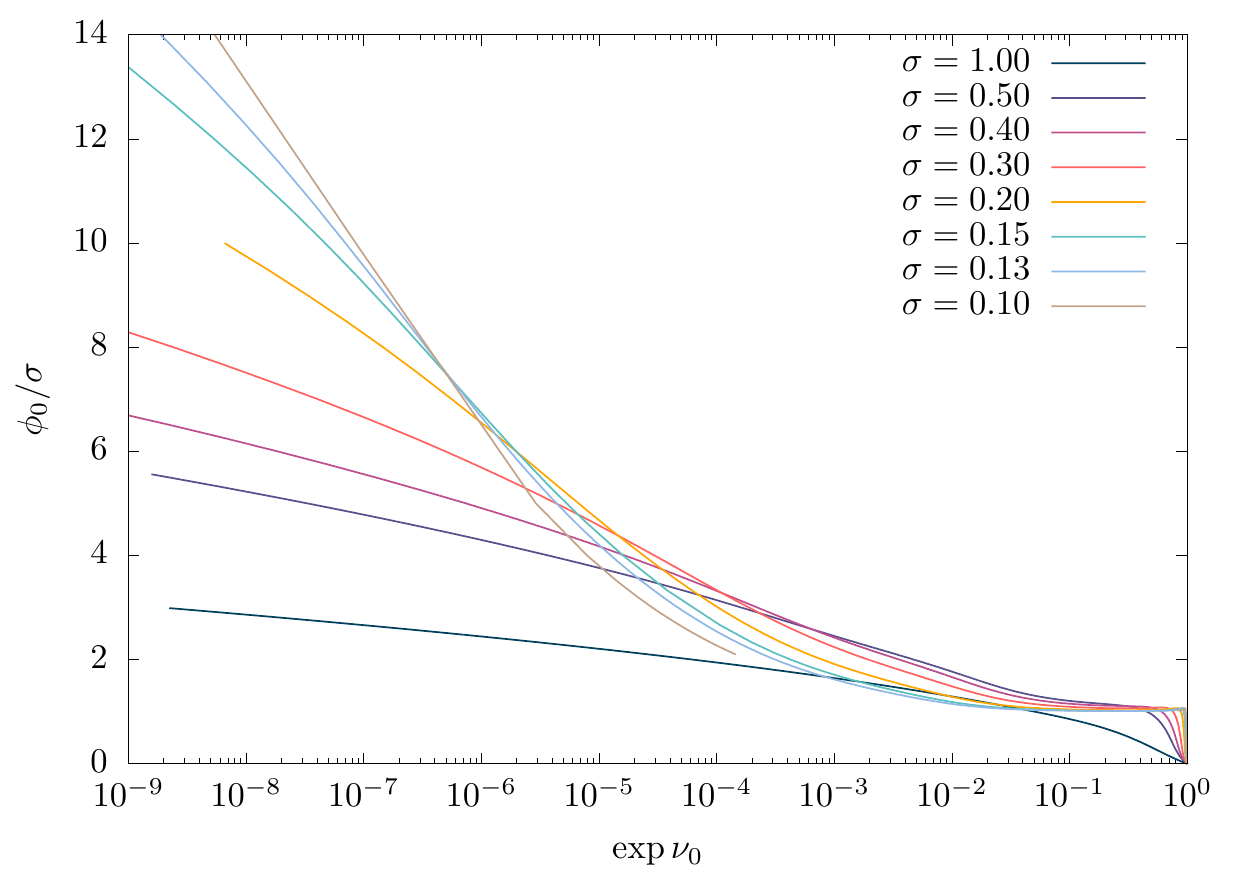}
\caption{\emph{Left}: Compactness vs. $\phi_0$, where we define $\mathcal{C}_5=M/R_5$. The dashed parts correspond to solutions containing lightrings, which can occur for $\mathcal{C}_5<1/3$, since they can dwell in the interior domain. The highest compactness we find from all of this solutions happens for $\sigma=0.13$, and not for $\sigma=0.1$ which does not present solutions in this specific region.
\emph{Right:} The central value of the field is displayed as a function of the central value of the metric function $\nu$. This value can actually parametrize the solutions as it uniquely determines them for each set. We see that asymptotically, as $\phi_0$ grows indefinitely, $g_{tt}$ tends to zero at the core.}
\label{Fig:3}
\end{figure}

In the left panel of Fig. \ref{Fig:3} we display the Kretschmann compactness (defined as $\mathcal{C}_5=M/R_5$) of the solutions with respect to $\phi_0/\sigma$. In all cases, the solution with the maximum compactness occur at the minimum value of $\omega_s$, still near $\phi_0=\sigma$. As expected, smaller $\sigma$ generally result in greater maximum compactness, but for $\sigma=0.1$ this fails to be true as the gap between the branches encompasses the region where the solution of maximum compactness should reside. The solid parts of each curve contain no lightrings while the dashed ones possess two. The filled area on the top of the plot encompasses the whole domain where $\mathcal{C}_5>1/3$. For any sensible definition of compactness, any solution that lies in this region should contain lightrings. Nevertheless, it is always possible to have lightrings in configurations of smaller compactness as those can reside in the stars interior. The maximum compactness $\mathcal{C}_5$ for each set is $\{0.321, 0.339, 0.334, 0.311, 0.260, 0.206, 0.160, 0.093 \}$ with increasing $\sigma$. This is in agreement with the result reported in \cite{Bo_kovi__2022} which has a lowest $\sigma$ value slightly higher than our $\sigma=0.13$, and a less stringent definition for the effective surface radius (one that encompasses $99\%$ of the mass).
In the right panel we present the central value of the scalar field as a function of the central value of $-g_{tt}$. While $\phi_0$ is not well suited to parametrize the solution, $\nu_0$ is. Its value uniquely determine the solution within each set defined by $\sigma$. As we mentioned previously, solutions continue to exist for ever-growing $\phi_0$, and we see from the figure that as we advance in this direction, $g_{tt}$ at the center approaches zero asymptotically. We remark that these solutions lie on the unstable branch and should probably not exist in nature and that the central value of $g_{rr}$ is set to unity in all cases via the boundary conditions. Even in the extreme case where $g_{tt}$ goes to zero at the center, that would not represent a black hole as there would be no null hypersurface. In fact, hairy black holes only exist for the given theory if the synchronization condition is satisfied , i.e. $\Omega_H=-\omega_s/m$ \cite{PhysRevLett.112.221101}, where $\Omega_H$ is the event horizon's angular velocity. Since we are restricting our analysis to static spacetimes, the condition cannot be satisfied for a nontrivial scalar field. Nevertheless, we remark that in the rotating case (where the curve is parametrized by $\partial_r^m\phi_0$), the solitonic curve approaches asymptotically the extremal hairy black holes curve \cite{PhysRevLett.112.221101}. Turning our attention back to the system at hand, if $\exp\nu_0\rightarrow 0$ as $\phi_0$ grows indefinitely, asymptotically one should get a naked singularity instead of a hole, as the scalar curvature at the origin
\be
\label{eq:R_0}
R_0=2\left[\left(1-\frac{\phi_0^2}{\sigma^2}\right)^2\mu^2-\omega_s^2 e^{-\nu_0}\right]\phi_0^2,
\ee
and the Kretschmann scalar
\be
\label{eq:K_0}
\mathcal{K}_0=\frac{4}{3}\phi_0^2\left[\left(1-\frac{\phi_0^2}{\sigma^2}\right)^2\mu^2R_0+5e^{-2\nu_0}\omega_s^4\phi_0^2\right]
\ee
both diverge. On the other hand, it is clear that in the idealized thin-wall regime ($\phi_0=\sigma$ and $\omega_s=0$) both quantities disappear at the center as the interior spacetime is flat.

\section{Compactness}
\label{sec:comp}
\begin{figure}[h!]
\centering
\includegraphics[width=0.48\linewidth]{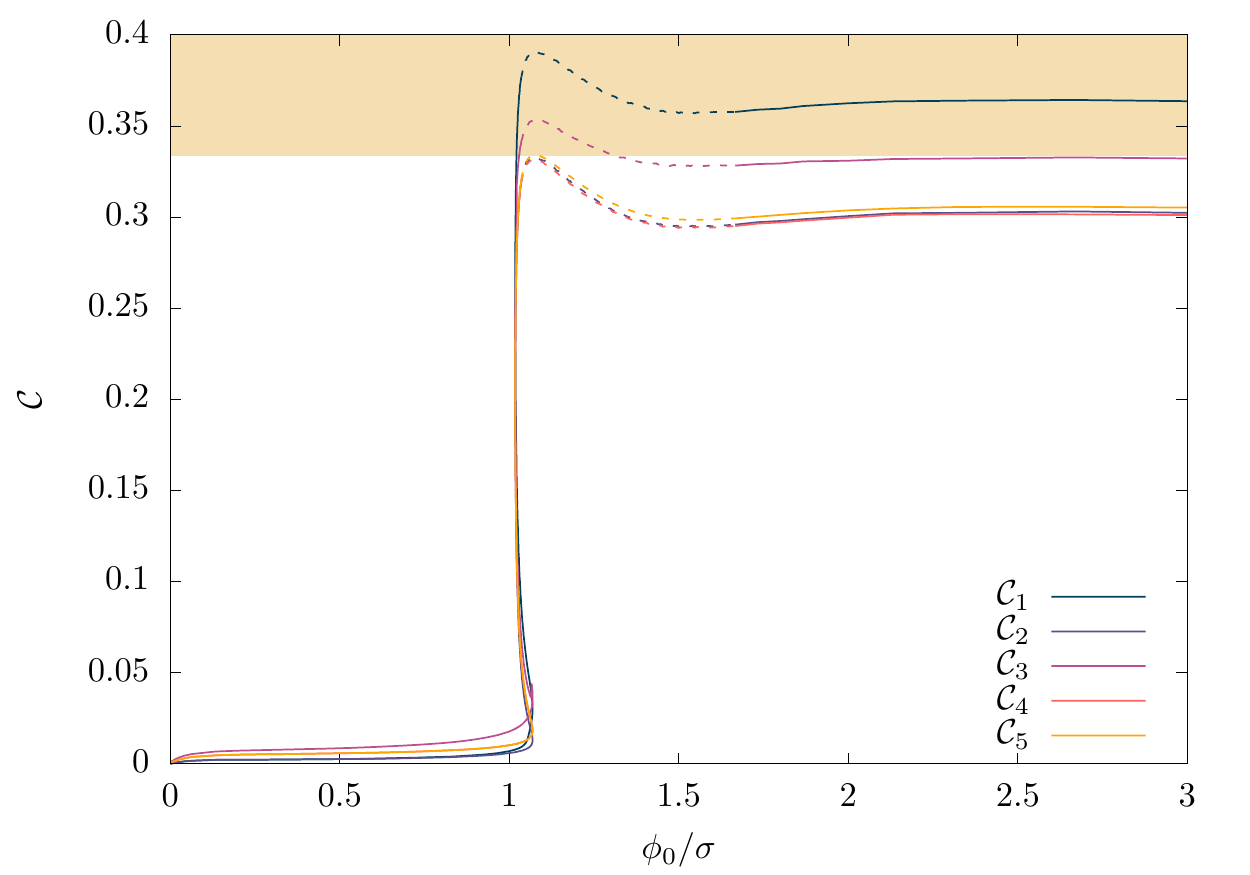}
\caption{The five differently defined compactness for the $\sigma=0.15$ set in terms of the scalar field's central value. Definitions $\mathcal{C}_2$, $\mathcal{C}_4$ and $\mathcal{C}_5$ fall very near one another at the regime of high compactness while $\mathcal{C}_1$ and $\mathcal{C}_3$ present much higher values. In fact, these latter two definitions are pathological as they allow solutions to feature $\mathcal{C}>1/3$ in the absence of lightrings. }
\label{Fig:hxC}
\end{figure}

In what follows we focus on one particular set of solutions, i.e. that of $\sigma=0.15$, and explore the different definitions of compactness given in \ref{Sub:Charge}. This is the set with largest value of $\sigma$ to achieve $\mathcal{C}>1/3$.  All five compactness for this set are plotted in Fig. \ref{Fig:hxC} with respect to $\phi_0/\sigma$. As previously, solutions lying in the solid parts contain no lightrings while those in the dashed possess two. We note that the definitions $C_2$, $C_4$ and $C_5$ fall near one another. Moreover, definitions $C_1$ and $C_3$ are pathological as within them a star might have a compactness greater than $1/3$ but contain no lightring. Overall, the definition of compactness is intuitively fuzzy as there is no definition of local mass in general relativity. In most static spacetimes this notion is more straightforward as there is a clear radial value which defines an exterior, Schwarzschild region. Because boson stars have no surface, this concept becomes ambiguous. Observationally, lightrings are probably the most important feature to characterize a threshold in any compactness definition. Objects with photonsphere cast shadows and are therefore good contenders for black hole mimickers. Yet, since this bosonic sector only interact gravitationally with ordinary matter, the existence of lightrings in the interior of the star yields even more confusion to the definition of a compactness. After all, if a noncompact star has a very dense core containing photonspheres and light and matter can move freely in there, we would perceive it as an exotic compact object.

\begin{figure}[h!]
\centering
\includegraphics[width=0.48\linewidth]{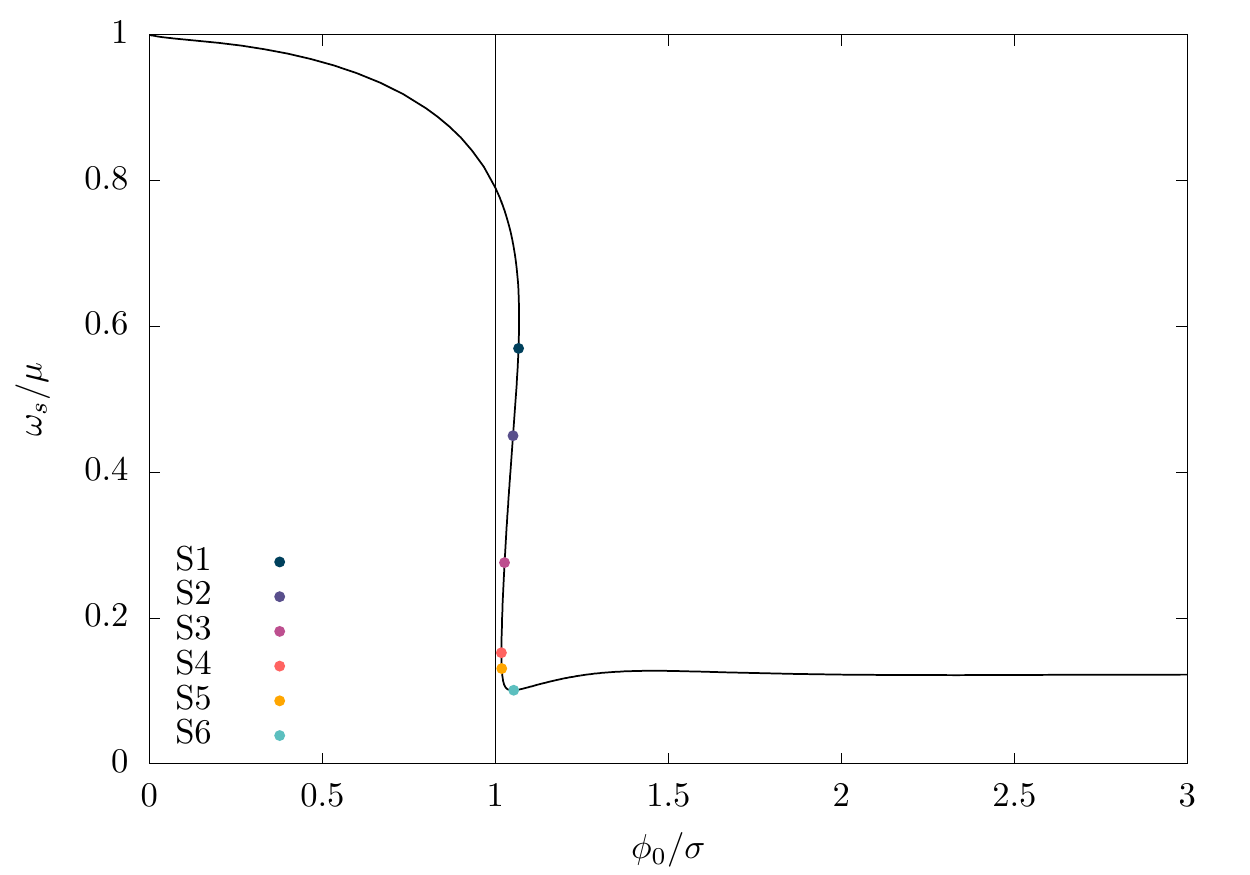}
\includegraphics[width=0.48\linewidth]{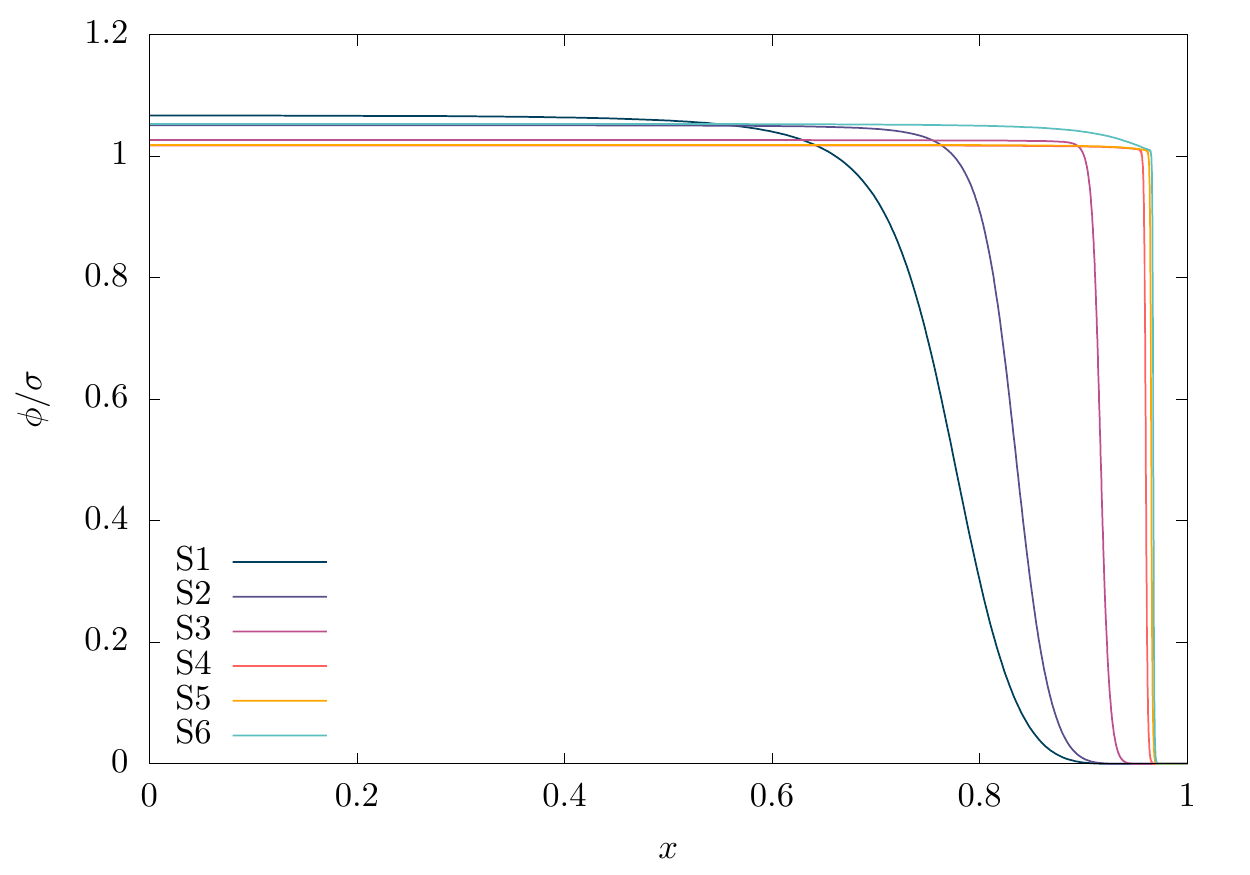}
\caption{Selected solutions in and around the thin-wall regime for $\sigma=0.15$. \emph{Left:} The field’s frequency against the central value of the field for the whole set with the six solutions highlighted. \emph{Right:} The scalar field's profile of each of those solutions with respect to the compactified coordinate $x$.}
\label{Fig:hxo_s=015}
\end{figure}

We select six configurations of stars in or near the thin-wall regime. Those are depicted in Fig. \ref{Fig:hxo_s=015}. The left panel displays the whole set in the $\phi_0/\sigma$ vs. $\omega_s/\mu$ diagram with the chosen solutions highlighted, while the right panel profiles these scalar fields with respect to the compactified coordinate $x$. Solutions $S_1$ and $S_2$ are still approaching the thin-wall regime and feature wider tails in the scalar field drop off. On the other hand, solution $S_6$ is departing from this region and while the field still drops off very steeply, it is a bit bulgy in the outer core. Such effect is enhanced as we move to the right in the left panel with increasing $\phi_0/\sigma$, as can be read off from the left panel of Fig. \ref{Fig:2}. 

The five values of compactness, corresponding to each definition, are displayed in Table \ref{Table:Compactness}. In any case, the solutions $S_1$ to $S_6$ go increasing in compactness. Overall, $C_2$ tends to be more conservative and provide the smallest values while $C_1$ yields the largest ones. As we move across the solutions, the differences between the definitions decrease and there is special good agreement between $C_2$, $C_4$ and $C_5$. The only solution featuring a compactness of $C>1/3$ according to at least one definition is $S_6$, which is the only one to containing lightrings indeed. We recall that lightrings occur at a point where the quantity $\delta_\text{lr}=2g_{tt}-rg'_{tt}$ becomes zero.
\begin{table}[h!]
\centering
\begin{tabular}{c || *{5}{c}}
\multicolumn{1}{c||}{}   & $C_1$ & $C_2$ & $C_3$ & $C_4$ & $C_5$ \\
\hline\hline
 $S_1$ & 0.036	& 0.016 & 0.043 & 0.021 & 0.021    \\
 $S_2$ & 0.057	& 0.029 & 0.046 & 0.034 & 0.034    \\
 $S_3$ & 0.121	& 0.081 & 0.098 & 0.085 & 0.085    \\
 $S_4$ & 0.258	& 0.204 & 0.222 & 0.206 & 0.207    \\
 $S_5$ & 0.301	& 0.245 & 0.263 & 0.246 & 0.247    \\
 $S_6$ & 0.388	& 0.330 & 0.350 & 0.329 & 0.332    \\ 
\hline\hline
\end{tabular}
\caption{Compactness of each selected solution.}
\label{Table:Compactness}
\end{table}

\begin{figure}[h!]
\centering
\includegraphics[width=0.48\linewidth]{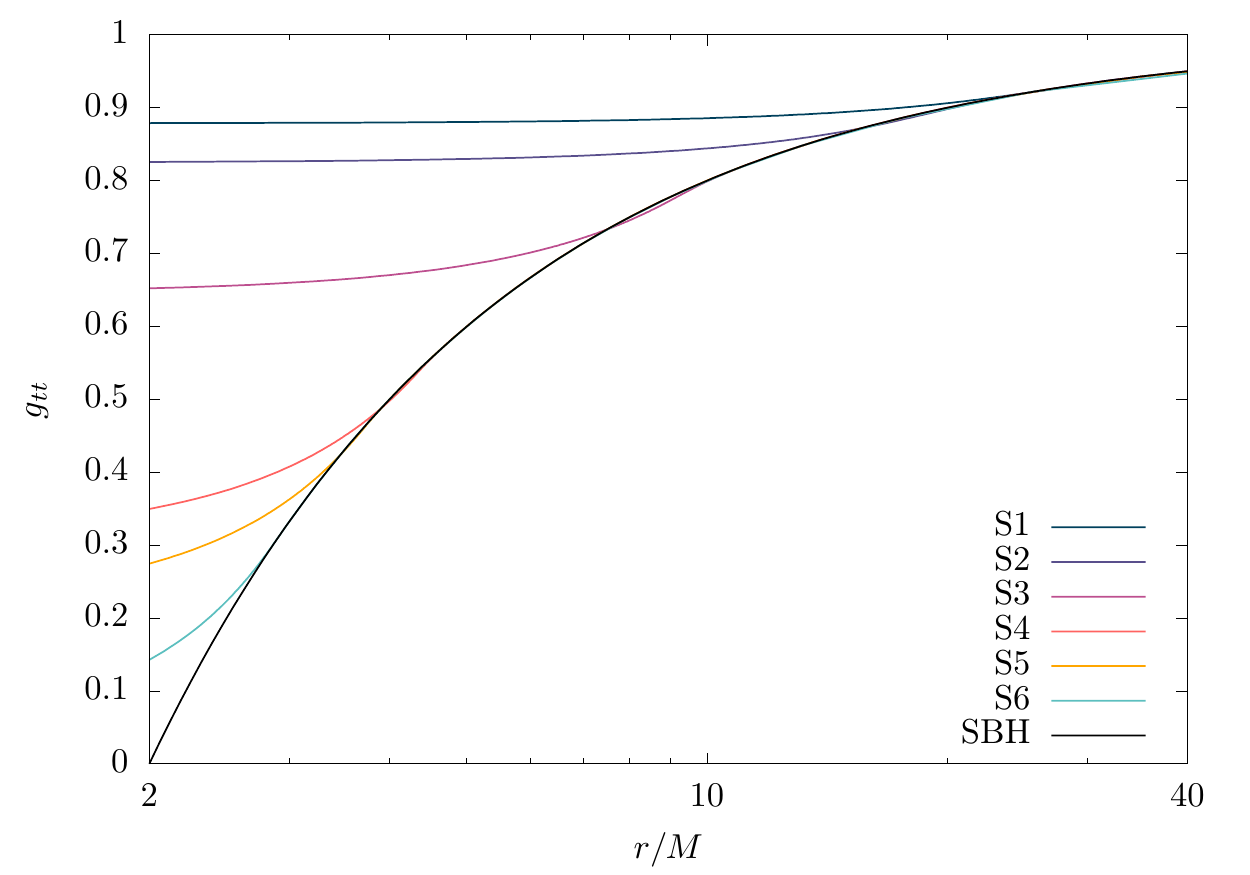}
\includegraphics[width=0.48\linewidth]{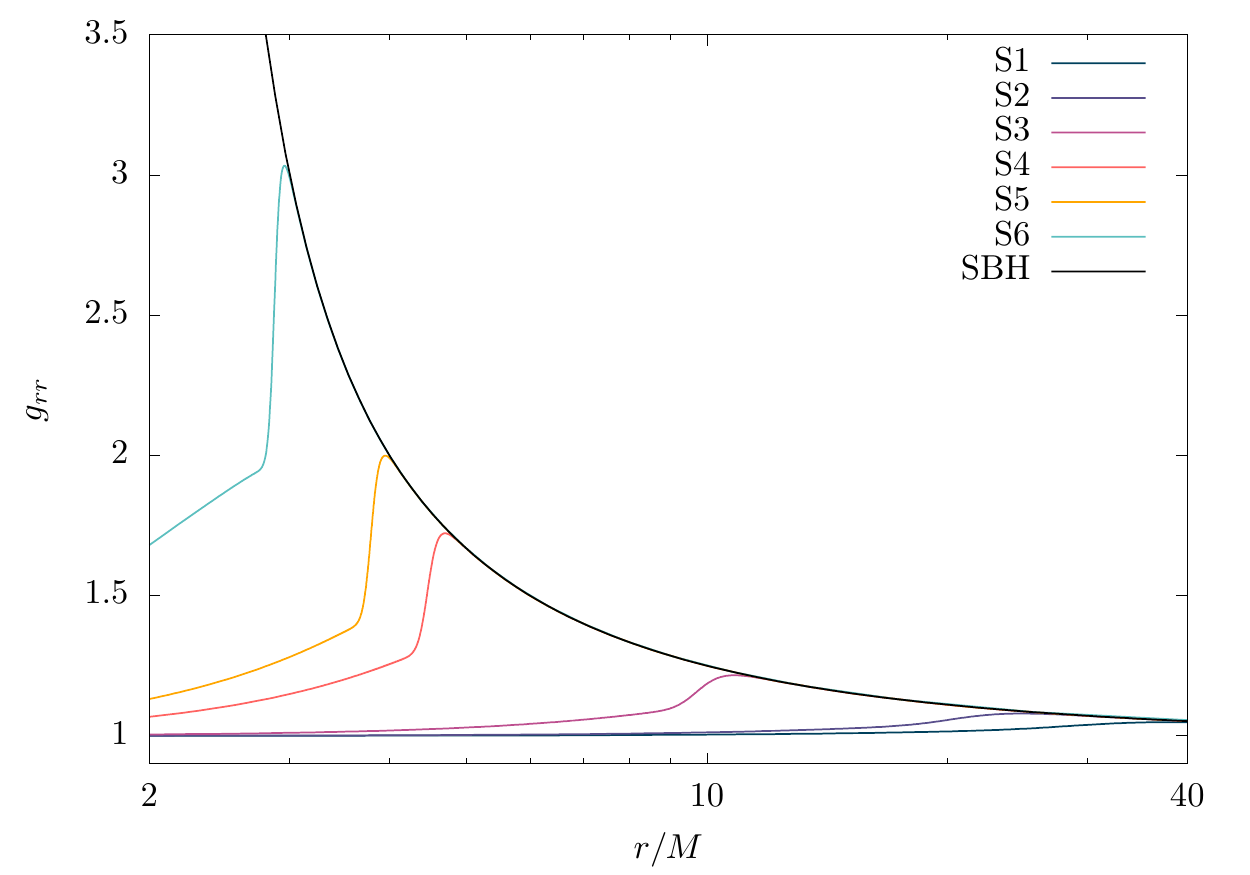}
\caption{Metric deviation from Schwarzschild (SBH) for the selected solutions with respect to the radial coordinate. \emph{Left:} $g_{tt}$ component. \emph{Right:} $g_{rr}$ component. }
\label{Fig:5}
\end{figure}

\begin{figure}[h!]
\centering
\includegraphics[width=0.48\linewidth]{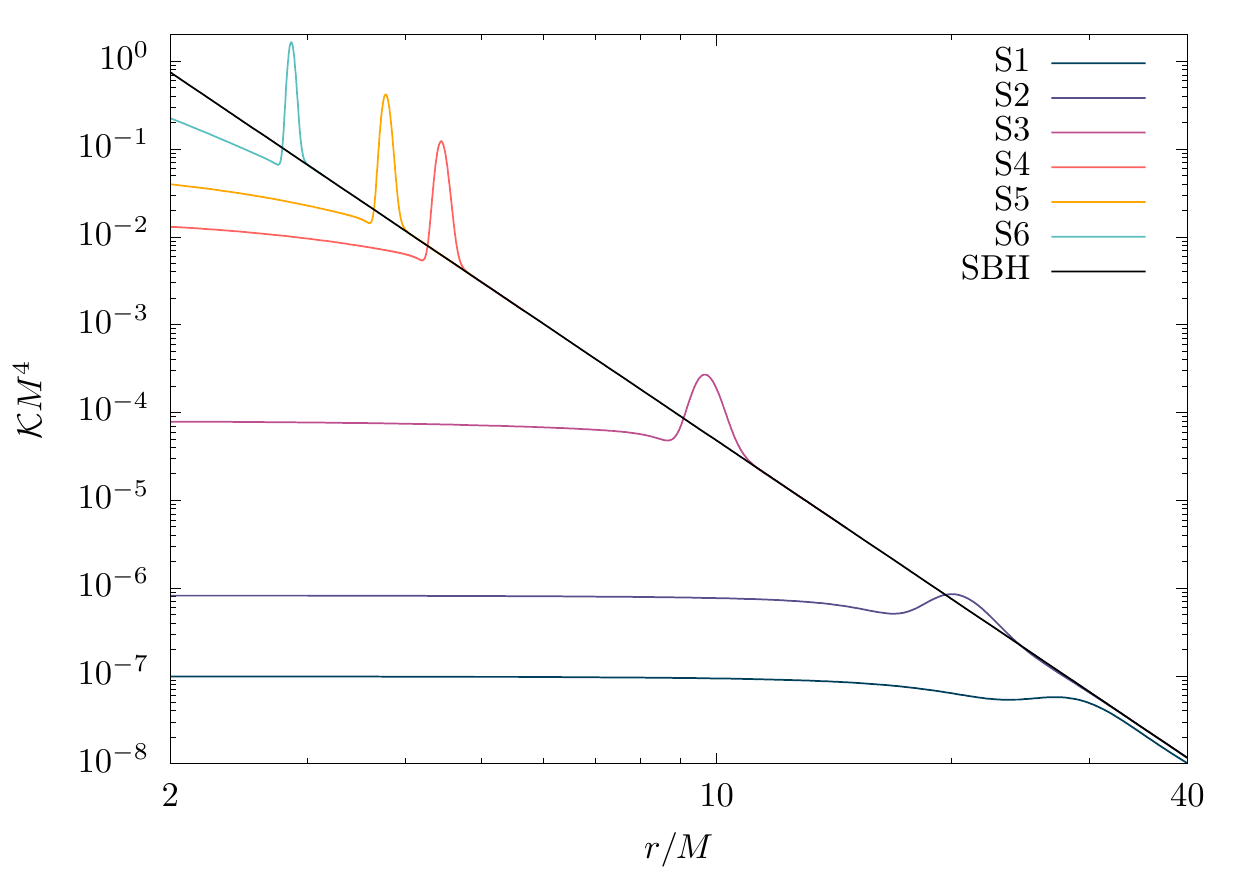}
\includegraphics[width=0.48\linewidth]{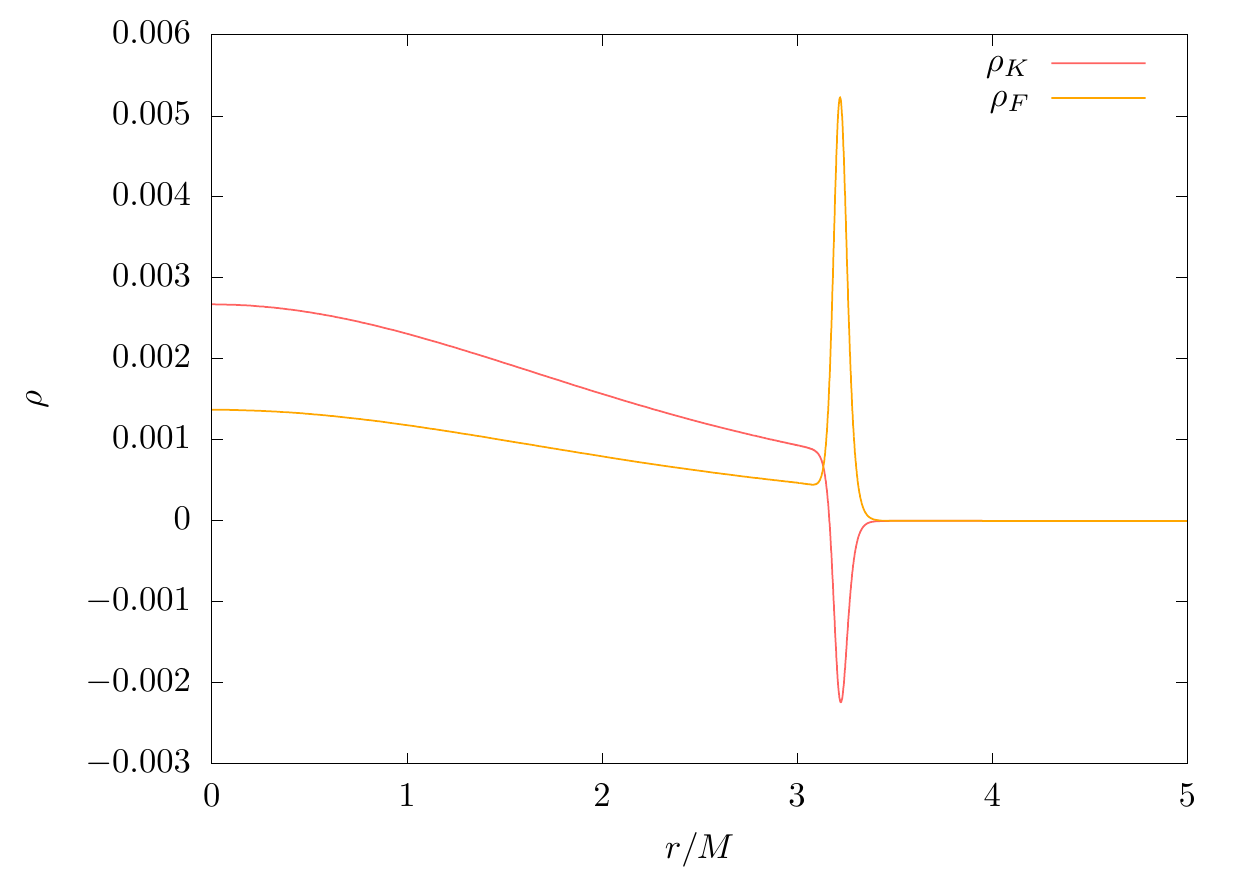}
\caption{\emph{Left:} Kretschmann scalar profile of each solution in contrast to that of a Schwarzschild black hole (SBH). \emph{Right:} The defined densities $\rho_K$ and $\rho_F$ for a typical solution. Since these systems violate the strong energy condition, $\rho_K<0$ in a certain region and is therefore not suitable to be used as a measure defining the compactness.}
\label{Fig:6}
\end{figure}

Definitions $C_3$, $C_4$ and $C_5$ are based on geometrical deviations from Schwarzschild. In Fig. \ref{Fig:5} we show the profile of the metric components of each selected solution with respect to the radial coordinate and contrast them with the Schwarzschild metric. In the left panel $g_{tt}$ is plotted against $r$, and it is clear that at the point of deviation from Schwarzschild, the slope of the curves are different. This is the reason why the first solutions to reach $C_3>1/3$ in Fig. \ref{Fig:hxC} contain no lightrings. The metric component $g_{rr}$ approaches Schwarzschild at a considerably higher radius than $g_{tt}$, as can be seen from the right panel, yielding a smaller compactness in $C_4$. Similarly, we display $C_5$ in the left panel of Fig. \ref{Fig:6}, which is defined with respect to the deviation in the Kretschmann scalar. This curvature invariant does not vary monotonically with the radius in the presence of matter. In particular, as it deviates from Schwarzschild with decreasing $r$, there is a local maximum followed by a local minimum in its profile. Thereon, the curvature might grow a small amount and remain nearly constant until the core (those solutions closer to the ideal thin-wall regime) or keep on growing all the way to the center of the star (and in many cases easily surpassing $0.75M^4$, that is its value at the event horizon for a Schwarzschild black hole).

In order to clarify the main difference between $C_1$ and $C_2$ and justify why $C_1$ is a bad choice of definition, we profile the densities $\rho_K$ and $\rho_F$ for a typical solution in the right panel of Fig. \ref{Fig:6}. The integrals of these densities are only equal in the interval $r\in[0,\infty)$. Furthermore, as these solutions violate the strong energy condition (SEC), $\rho_K$ is negative in some radial domain. Hence, by integrating it subsequently in finite intervals of $r$, the value of $0.999M$ is reached much earlier than when integrating $\rho_F$, as the integral yields values above $M$ before reaching the region where $\rho_K < 0$. In passing, let us mention that solutions that violate the SEC are much more abundant in solitonic theories than others. Thus $\mathcal{C}_1$ may still give a reasonable measure of compactness for noninteracting stars or BSs with a quartic potential. By analyzing $\rho_K$, we see it is negative when $2\omega_s^2\phi^2<e^\nu U$. Because the potential (\ref{potential}) has a local maximum between $\phi=0$ and $\phi=\sigma$, this inequality is realized around the point where the scalar field drops off. 

\section{Conclusions}
\label{sec:conclusions}

In this paper we investigated the existence domain of solitonic boson stars for several values of the parameter $\sigma$ which controls the solitonic potential \eqref{potential} and  defines the interacting theory. We have paid special attention to values of the parameters leading to solutions belonging to the so-called thin-wall regime, where the scalar field inside the solitonic star is nearly constant followed by a sharp drop that is exactly the thin-wall. Such solutions are particularly interesting because they can achieve very high compactness thus serving as an efficient black hole mimickers. Contrary to most of the previous studies, the solutions in the present paper were obtained numerically and self-consistently without any approximation.

As the limit $\sigma\rightarrow\infty$ corresponds to the free scalar field theory, by decreasing $\sigma$ the solution sets depart from that resembling mini boson stars and particular features of solitonic stars become more prominent. In particular, while for large $\sigma$ the solutions are uniquely identified via the central value of the scalar field $\phi_0$, for small $\sigma$ values there is a region in the parameter space near $\phi_0/\sigma\sim 1$ where three different solutions can be found. This region contains the solutions belonging to the thin-wall regime. The range in $\omega_s$, the field's natural frequency, that each solution set spans is inversely proportional to $\sigma$. Also, as $\sigma$ gets smaller, wider is the range in $\omega_s$ within the nonuniqueness region. Similarly to ground state boson stars with different interacting potentials, solutions can be found for continuously increasing values of $\phi_0$ where $\omega_s$ asymptotically oscillates around a fixed point. A parameter that can serve to uniquely determine each solution is the value of $g_{tt}$ at the origin, with which one can therefore parametrize the solution. As $\phi_0\rightarrow\infty$, $g_{tt}\rightarrow 0$, and the curvature invariants are singular at the center. Besides being unstable, if solutions in this limit existed they would not be black holes for the lack of the null hypersurface, but would characterize naked singularities instead.

Through the construction of several families of solutions each defined by the $\sigma$, we could show that as $\sigma$ decreases the closer the curves get to the line of $\phi_0/\sigma=1$ representing the thin-wall regime. Nevertheless, by expanding the equations around this point we showed that no solution exists exactly over this line, in this regime. Thus, the approximate treatment of this thin-wall regime that can be found in the literature, albeit helpful in mimicking configurations near  $\phi_0/\sigma=1$, does not correspond to real solutions. We also did not find solutions in this region for sufficiently small values of $\sigma$, and while this could simply be due to numerical difficulties, we cannot rule out the possibility that some of those solutions do not exist as they would approach the vertical $\phi_0/\sigma=1$ line ever more, to the point where eq. (\ref{AESp}) cannot be satisfied anymore.

An important part of the paper is the discussion on the possible definitions of compactness since as it turns out, some of the standard definitions might be misleading for solitonic BSs, or more generally stars whose scalar field theory accounts for the existence of a false vacuum. We have analyzed five different radius (and therefore compactness) definitions. The first two are the radii that contains 99\% of the star's mass, calulated differently. The remaining three are based on geometrical deviation from Schwarzschild. We showed that two of them are bad measure, namely $\mathcal{C}_1$ and $\mathcal{C}_3$ defined at the end of section \ref{Sub:Charge}. The first one $\mathcal{C}_1$ uses the radius that contains $99.9\%$ of the mass when integrating the Komar expression. This raises issues as the integrand cannot be taken to be a local density measurement, and because these configurations violate the strong energy condition. Hence, the integrand in the Kommar mass integral can be negative in some radial range and the calculated radius turns out to be much smaller than any other effective one seems to be. The second problematic compactness definition $\mathcal{C}_3$  uses the smallest radial value for which the $g_{tt}$ component of the metric deviates less than $0.1\%$ from Schwarzschild. But because the slope of $g_{tt}$ at this point is much different than that of a Schwarzschild metric, the radius it yields falls also short of an effective one. In turn, these definitions allow for configurations with a compactness $\mathcal{C}>1/3$ but featuring no lightrings. In the other three definitions $\mathcal{C}_2$, $\mathcal{C}_4$ and $\mathcal{C}_5$, we employed, presented no such issues and yielded similar results among themselves. The maximum compactness (defined in terms of the Kretschmann invariant) we obtained is of $\mathcal{C}_5=0.339$, found for the $\sigma=0.13$ set. In a future work, we intend to explore the properties of these solutions in astrophysical phenomenology and investigate what imprints could there be to tell one compact ($\mathcal{C}>1/3$) from another, and from a black hole.

\section*{Acknowledgements}
LC and DD acknowledge financial support via an Emmy Noether Research Group funded by the German Research Foundation (DFG) under grant no. DO 1771/1-1. Networking support by the COST Actions  CA16104 and CA16214 is also gratefully acknowledged.

\appendix

\bibliographystyle{ieeetr}
\bibliography{biblio}

\begin{thebibliography}{10}

\bibitem{PhysRev.172.1331}
D.~J. Kaup, ``Klein-gordon geon,'' {\em Phys. Rev.}, vol.~172, pp.~1331--1342,
  Aug 1968.

\bibitem{PhysRev.187.1767}
R.~RUFFINI and S.~BONAZZOLA, ``Systems of self-gravitating particles in general
  relativity and the concept of an equation of state,'' {\em Phys. Rev.},
  vol.~187, pp.~1767--1783, Nov 1969.

\bibitem{Coleman:1985ki}
S.~R. Coleman, ``{Q-balls},'' {\em Nucl. Phys. B}, vol.~262, no.~2, p.~263,
  1985.
\newblock [Addendum: Nucl.Phys.B 269, 744 (1986)].

\bibitem{PhysRevD.66.085003}
M.~S. Volkov and E.~W\"ohnert, ``Spinning q-balls,'' {\em Phys. Rev. D},
  vol.~66, p.~085003, Oct 2002.

\bibitem{PhysRevD.72.064002}
B.~Kleihaus, J.~Kunz, and M.~List, ``Rotating boson stars and $q$-balls,'' {\em
  Phys. Rev. D}, vol.~72, p.~064002, Sep 2005.

\bibitem{PhysRevD.77.064025}
B.~Kleihaus, J.~Kunz, M.~List, and I.~Schaffer, ``Rotating boson stars and
  $q$-balls. ii. negative parity and ergoregions,'' {\em Phys. Rev. D},
  vol.~77, p.~064025, Mar 2008.

\bibitem{PhysRevD.83.044027}
T.~Tamaki and N.~Sakai, ``How does gravity save or kill $q$-balls?,'' {\em
  Phys. Rev. D}, vol.~83, p.~044027, Feb 2011.

\bibitem{PhysRevD.85.024045}
B.~Kleihaus, J.~Kunz, and S.~Schneider, ``Stable phases of boson stars,'' {\em
  Phys. Rev. D}, vol.~85, p.~024045, Jan 2012.

\bibitem{PhysRevD.96.084066}
L.~G. Collodel, B.~Kleihaus, and J.~Kunz, ``Excited boson stars,'' {\em Phys.
  Rev. D}, vol.~96, p.~084066, Oct 2017.

\bibitem{PhysRevD.99.104076}
L.~G. Collodel, B.~Kleihaus, and J.~Kunz, ``Structure of rotating charged boson
  stars,'' {\em Phys. Rev. D}, vol.~99, p.~104076, May 2019.

\bibitem{PhysRevD.13.2739}
R.~Friedberg, T.~D. Lee, and A.~Sirlin, ``Class of scalar-field soliton
  solutions in three space dimensions,'' {\em Phys. Rev. D}, vol.~13,
  pp.~2739--2761, May 1976.

\bibitem{PhysRevD.35.3637}
T.~D. Lee, ``Soliton stars and the critical masses of black holes,'' {\em Phys.
  Rev. D}, vol.~35, pp.~3637--3639, Jun 1987.

\bibitem{Friedberg:1986tq}
R.~Friedberg, T.~D. Lee, and Y.~Pang, ``{Scalar Soliton Stars and Black
  Holes},'' {\em Phys. Rev. D}, vol.~35, p.~3658, 1987.

\bibitem{Liebling:2012fv}
S.~L. Liebling and C.~Palenzuela, ``{Dynamical Boson Stars},'' {\em Living Rev.
  Rel.}, vol.~15, p.~6, 2012.

\bibitem{PhysRevLett.112.221101}
C.~A.~R. Herdeiro and E.~Radu, ``Kerr black holes with scalar hair,'' {\em
  Phys. Rev. Lett.}, vol.~112, p.~221101, Jun 2014.

\bibitem{PhysRevD.80.084023}
F.~S. Guzm\'an and J.~M. Rueda-Becerril, ``Spherical boson stars as black hole
  mimickers,'' {\em Phys. Rev. D}, vol.~80, p.~084023, Oct 2009.

\bibitem{Cao_2016}
Z.~Cao, A.~C{\'{a}}rdenas-Avenda{\~{n}}o, M.~Zhou, C.~Bambi, C.~A. Herdeiro,
  and E.~Radu, ``Iron k$\alpha$ line of boson stars,'' {\em Journal of
  Cosmology and Astroparticle Physics}, vol.~2016, pp.~003--003, oct 2016.

\bibitem{Meliani_2017}
Z.~Meliani, F.~Casse, P.~Grandcl{\'{e}}ment, E.~Gourgoulhon, and F.~Dauvergne,
  ``On tidal disruption of clouds and disk formation near boson stars,'' {\em
  Classical and Quantum Gravity}, vol.~34, p.~225003, oct 2017.

\bibitem{Teodoro_2021}
M.~C. Teodoro, L.~G. Collodel, and J.~Kunz, ``Retrograde polish doughnuts
  around boson stars,'' {\em Journal of Cosmology and Astroparticle Physics},
  vol.~2021, p.~063, mar 2021.

\bibitem{PhysRevD.103.104064}
M.~C. Teodoro, L.~G. Collodel, and J.~Kunz, ``Tidal effects in the motion of
  gas clouds around boson stars,'' {\em Phys. Rev. D}, vol.~103, p.~104064, May
  2021.

\bibitem{Collodel_2021}
L.~G. Collodel, D.~D. Doneva, and S.~S. Yazadjiev, ``Circular orbit structure
  and thin accretion disks around kerr black holes with scalar hair,'' {\em The
  Astrophysical Journal}, vol.~910, p.~52, mar 2021.

\bibitem{Teodoro:2021ezj}
M.~C. Teodoro, L.~G. Collodel, D.~Doneva, J.~Kunz, P.~Nedkova, and
  S.~Yazadjiev, ``{Thick toroidal configurations around scalarized Kerr black
  holes},'' {\em Phys. Rev. D}, vol.~104, no.~12, p.~124047, 2021.

\bibitem{Vincent_2016}
F.~H. Vincent, Z.~Meliani, P.~Grandcl{\'{e}}ment, E.~Gourgoulhon, and
  O.~Straub, ``Imaging a boson star at the galactic center,'' {\em Classical
  and Quantum Gravity}, vol.~33, p.~105015, apr 2016.

\bibitem{Olivares:2018abq}
H.~Olivares, Z.~Younsi, C.~M. Fromm, M.~De~Laurentis, O.~Porth, Y.~Mizuno,
  H.~Falcke, M.~Kramer, and L.~Rezzolla, ``{How to tell an accreting boson star
  from a black hole},'' {\em Mon. Not. Roy. Astron. Soc.}, vol.~497, no.~1,
  pp.~521--535, 2020.

\bibitem{PhysRevD.71.044015}
M.~Kesden, J.~Gair, and M.~Kamionkowski, ``Gravitational-wave signature of an
  inspiral into a supermassive horizonless object,'' {\em Phys. Rev. D},
  vol.~71, p.~044015, Feb 2005.

\bibitem{PhysRevD.88.064046}
C.~F.~B. Macedo, P.~Pani, V.~Cardoso, and L.~C.~B. Crispino, ``Astrophysical
  signatures of boson stars: Quasinormal modes and inspiral resonances,'' {\em
  Phys. Rev. D}, vol.~88, p.~064046, Sep 2013.

\bibitem{PhysRevD.94.084031}
V.~Cardoso, S.~Hopper, C.~F.~B. Macedo, C.~Palenzuela, and P.~Pani,
  ``Gravitational-wave signatures of exotic compact objects and of quantum
  corrections at the horizon scale,'' {\em Phys. Rev. D}, vol.~94, p.~084031,
  Oct 2016.

\bibitem{PhysRevD.95.124005}
M.~Bezares, C.~Palenzuela, and C.~Bona, ``Final fate of compact boson star
  mergers,'' {\em Phys. Rev. D}, vol.~95, p.~124005, Jun 2017.

\bibitem{PhysRevD.96.104058}
C.~Palenzuela, P.~Pani, M.~Bezares, V.~Cardoso, L.~Lehner, and S.~Liebling,
  ``Gravitational wave signatures of highly compact boson star binaries,'' {\em
  Phys. Rev. D}, vol.~96, p.~104058, Nov 2017.

\bibitem{Cardoso_2022}
V.~Cardoso, C.~F.~B. Macedo, K.~ichi Maeda, and H.~Okawa, ``{ECO}-spotting:
  looking for extremely compact objects with bosonic fields,'' {\em Classical
  and Quantum Gravity}, vol.~39, p.~034001, jan 2022.

\bibitem{PhysRev.116.1027}
H.~A. Buchdahl, ``General relativistic fluid spheres,'' {\em Phys. Rev.},
  vol.~116, pp.~1027--1034, Nov 1959.

\bibitem{Urbano_2019}
A.~Urbano and H.~Veermäe, ``On gravitational echoes from ultracompact exotic
  stars,'' {\em Journal of Cosmology and Astroparticle Physics}, vol.~2019,
  pp.~011--011, apr 2019.

\bibitem{Bo_kovi__2022}
M.~Bo{\v{s}}kovi{\'{c}} and E.~Barausse, ``Soliton boson stars, q-balls and the
  causal buchdahl bound,'' {\em Journal of Cosmology and Astroparticle
  Physics}, vol.~2022, p.~032, feb 2022.

\bibitem{PhysRevD.92.124061}
Y.~Brihaye, A.~Cisterna, B.~Hartmann, and G.~Luchini, ``From topological to
  nontopological solitons: Kinks, domain walls, and $q$-balls in a scalar field
  model with a nontrivial vacuum manifold,'' {\em Phys. Rev. D}, vol.~92,
  p.~124061, Dec 2015.

\bibitem{10.1145/355945.355951}
U.~Ascher, J.~Christiansen, and R.~D. Russell, ``Algorithm 569: Colsys:
  Collocation software for boundary-value odes [d2],'' {\em ACM Trans. Math.
  Softw.}, vol.~7, p.~223–229, jun 1981.

\end{thebibliography}
\end{document}